\documentclass[journal]{IEEEtran}

\usepackage{cite}
\usepackage{csquotes}
\usepackage[T1]{fontenc}
\usepackage{tabularx}
\usepackage{booktabs}
\usepackage{multirow}
\usepackage{caption}
\usepackage{subcaption}
\usepackage{algpseudocode}
\usepackage{algorithm}
\usepackage{comment}
\usepackage[scaled]{helvet}
\usepackage{booktabs}
\usepackage{hyperref}

\usepackage{xspace}

\newcommand{\latinphrase}[1]{\textit{#1}} 
\newcommand{\etal}{\latinphrase{et~al.}\xspace}
\newcommand{\ie}{\latinphrase{i.e.}\xspace}

\newcommand{\etc}{\latinphrase{etc.}\xspace}

\usepackage{seqsplit}
\usepackage[dvipsnames]{xcolor}

\usepackage{tabularx}

\newcolumntype{L}[1]{>{\raggedright\arraybackslash}m{#1}}
\newcolumntype{C}[1]{>{\centering\arraybackslash}m{#1}}
\newcolumntype{R}[1]{>{\raggedleft\arraybackslash}m{#1}}

\usepackage{cite}
\usepackage{dblfloatfix} 
\usepackage{multirow}
\ifCLASSINFOpdf
  \usepackage[pdftex]{graphicx}
\else
\fi

\usepackage[cmex10]{amsmath}
\usepackage{url}
\begin{document}

\title{Vehicle and License Plate Recognition \\with Novel Dataset for Toll Collection}

\author{Muhammad~Usama,
        Hafeez~Anwar,
        Abbas~Anwar,
        Saeed~Anwar,

		\thanks{M. Usama, H. Anwar, and A. Anwar are with The Department of Electrical and Computer Engineering, COMSATS University Islamabad, Attock Campus, Pakistan
        e-mail: hafeez.anwar@cuiatk.edu.pk (Corresponding Author)}
        \thanks{S. Anwar is with the Commonwealth Scientific and Industrial Research Organisation (CSIRO), the Australian National University (ANU), and the University of Technology Sydney (UTS), Australia}
			
		}

\maketitle
\begin{abstract}
 We propose an automatic framework for toll collection, consisting of three steps: vehicle type recognition, license plate localization, and reading. However, each of the three steps becomes non-trivial due to image variations caused by several factors. The traditional vehicle decorations on the front cause variations among vehicles of the same type. These decorations make license plate localization and recognition difficult due to severe background clutter and partial occlusions. Likewise, on most vehicles, specifically trucks, the position of the license plate is not consistent. Lastly, for license plate reading, the variations are induced by non-uniform font styles, sizes, and partially occluded letters and numbers. Our proposed framework takes advantage of both data availability and performance evaluation of the backbone deep learning architectures. We gather a novel dataset, \emph{Diverse Vehicle and License Plates Dataset (DVLPD)}, consisting of 10k images belonging to six vehicle types. Each image is then manually annotated for vehicle type, license plate, and its characters and digits. For each of the three tasks, we evaluate You Only Look Once (YOLO)v2, YOLOv3, YOLOv4, and FasterRCNN. For real-time implementation on a Raspberry Pi, we evaluate the lighter versions of YOLO named Tiny YOLOv3 and Tiny YOLOv4. 
The best Mean Average Precision (mAP@0.5) of 98.8\% for vehicle type recognition, 98.5\% for license plate detection, and 98.3\% for license plate reading is achieved by YOLOv4, while its lighter version \ie, Tiny YOLOv4 obtained a mAP of 97.1\%, 97.4\%, and 93.7\% on vehicle type recognition, license plate detection, and license plate reading, respectively. The dataset and the training codes are available at \href{https://github.com/usama-x930/VT-LPR}{https://github.com/usama-x930/VT-LPR} 
\end{abstract}


\IEEEpeerreviewmaketitle


\section{Introduction}

\begin{figure}[t!]
    \centering
		\resizebox{\columnwidth}{!}{%
    \begin{tabular}[t]{cc}
    
					\begin{tabular}{c}
						
								\begin{subfigure}[t]{0.4\textwidth}
									\centering
									\includegraphics[width=1\textwidth]{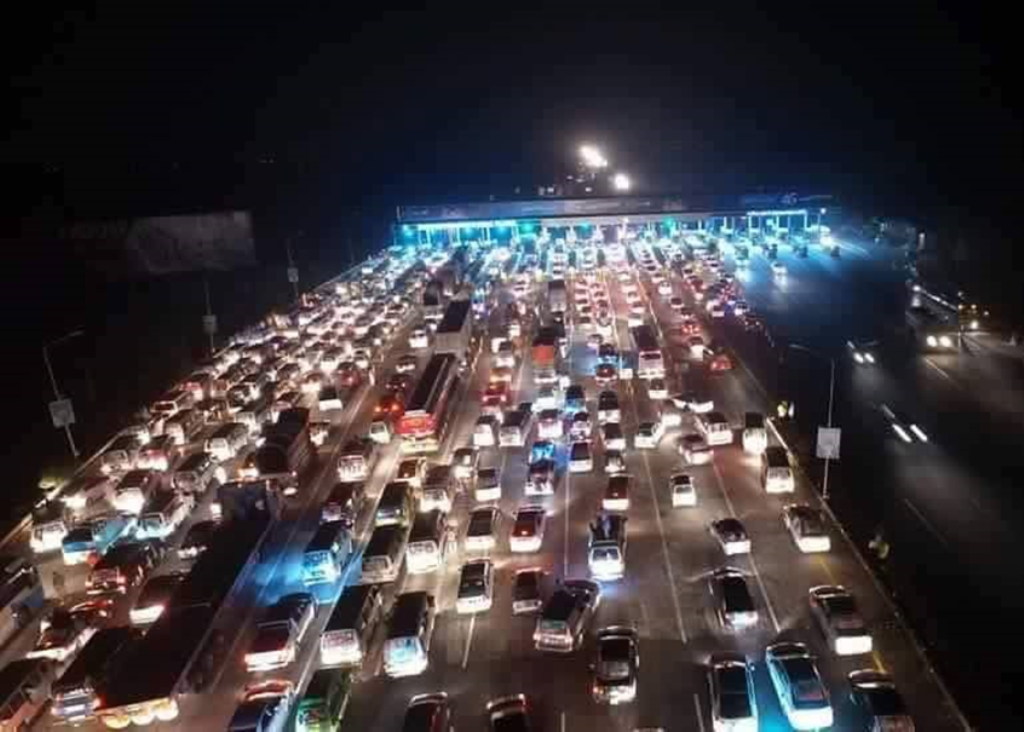}
									
								\end{subfigure}\\
								\begin{subfigure}[t]{0.4\textwidth}
									\centering
									\includegraphics[width=1\textwidth]{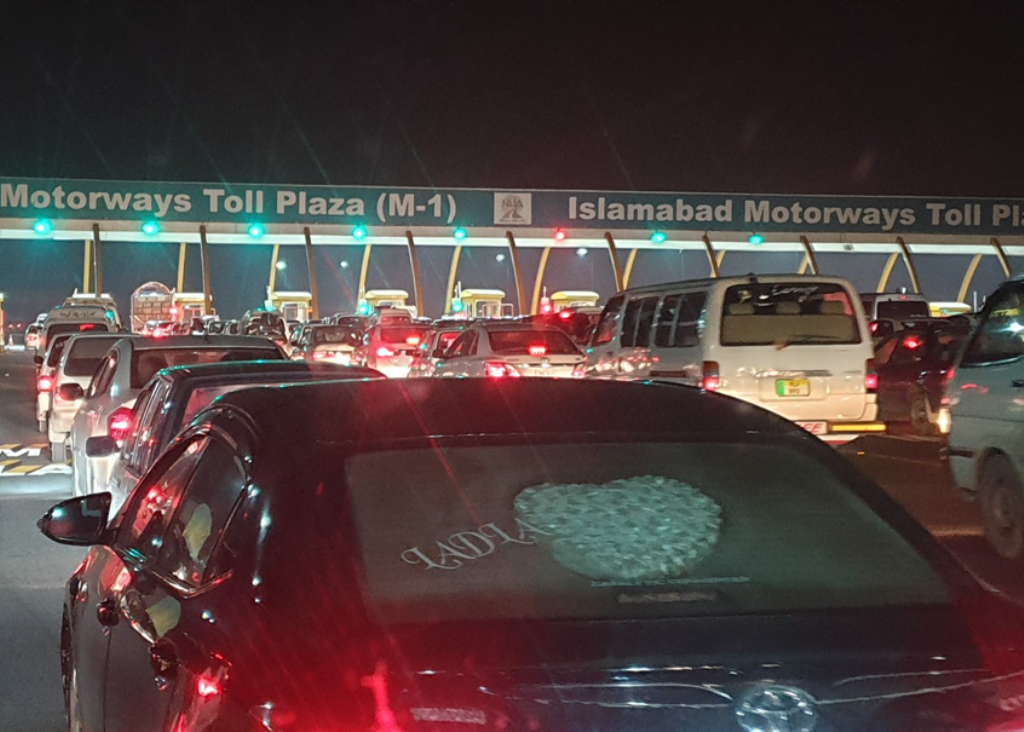}
									
							\end{subfigure}\\
							\begin{subfigure}[t]{0.4\textwidth}
									\centering
									\includegraphics[width=1\textwidth]{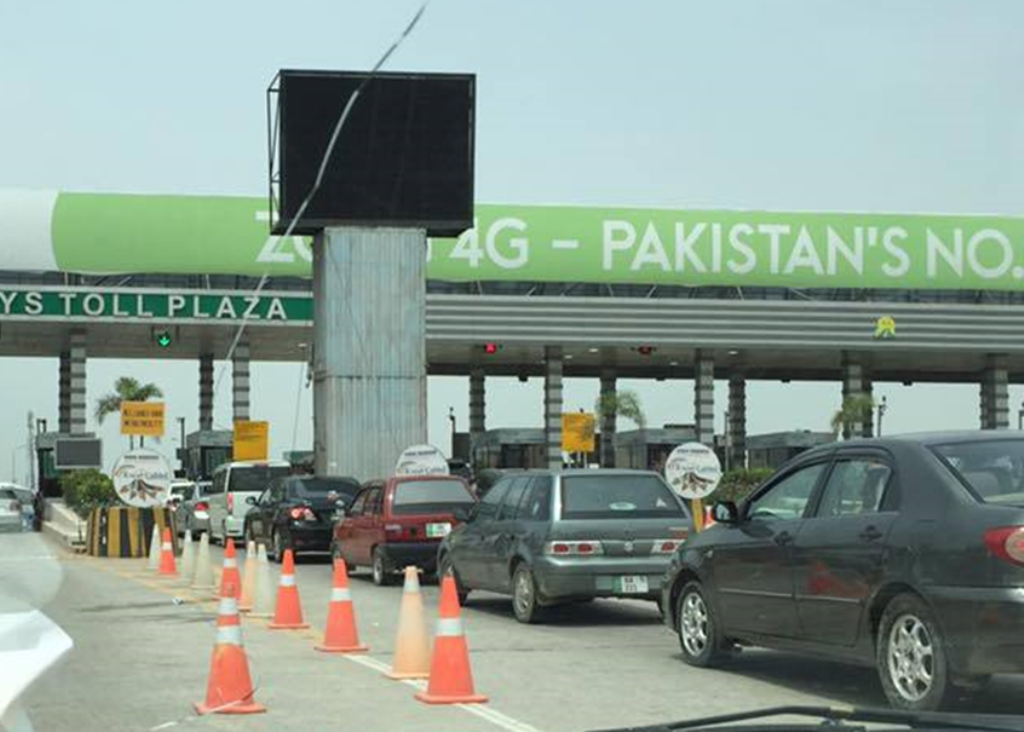}
									
							\end{subfigure}\\
							
					\end{tabular}
			& 
		  \begin{tabular}{c}
						
								\begin{subfigure}[t]{0.4\textwidth}
									\centering
									\includegraphics[width=1\textwidth]{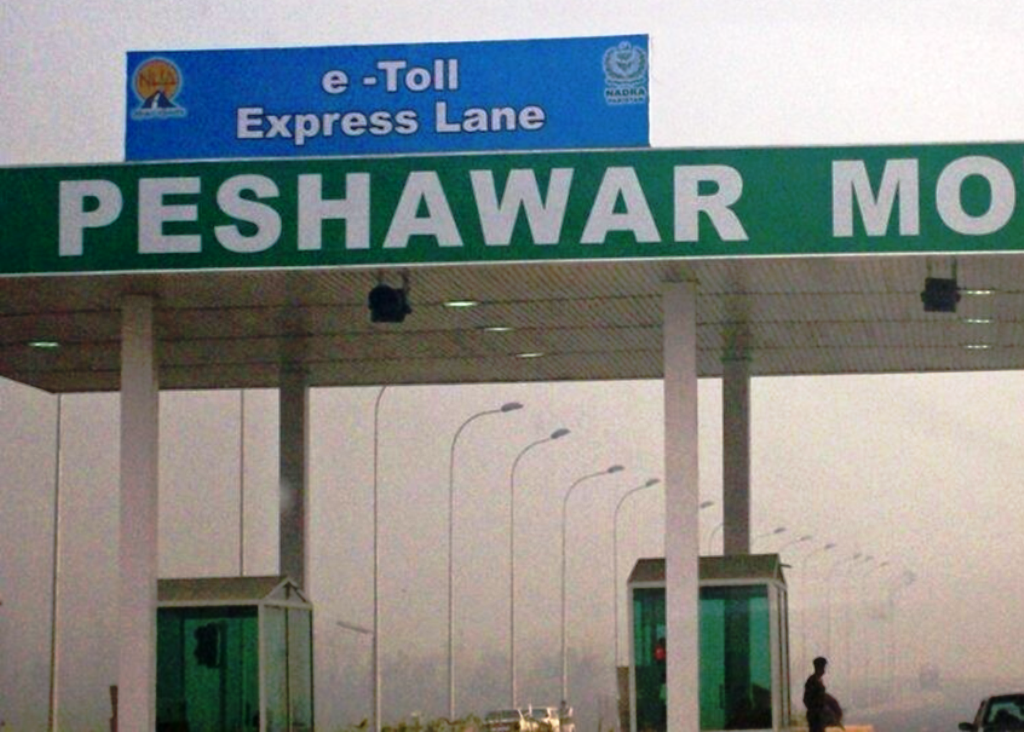}
									
								\end{subfigure}\\
								\begin{subfigure}[t]{0.4\textwidth}
									\centering
									\includegraphics[width=1\textwidth]{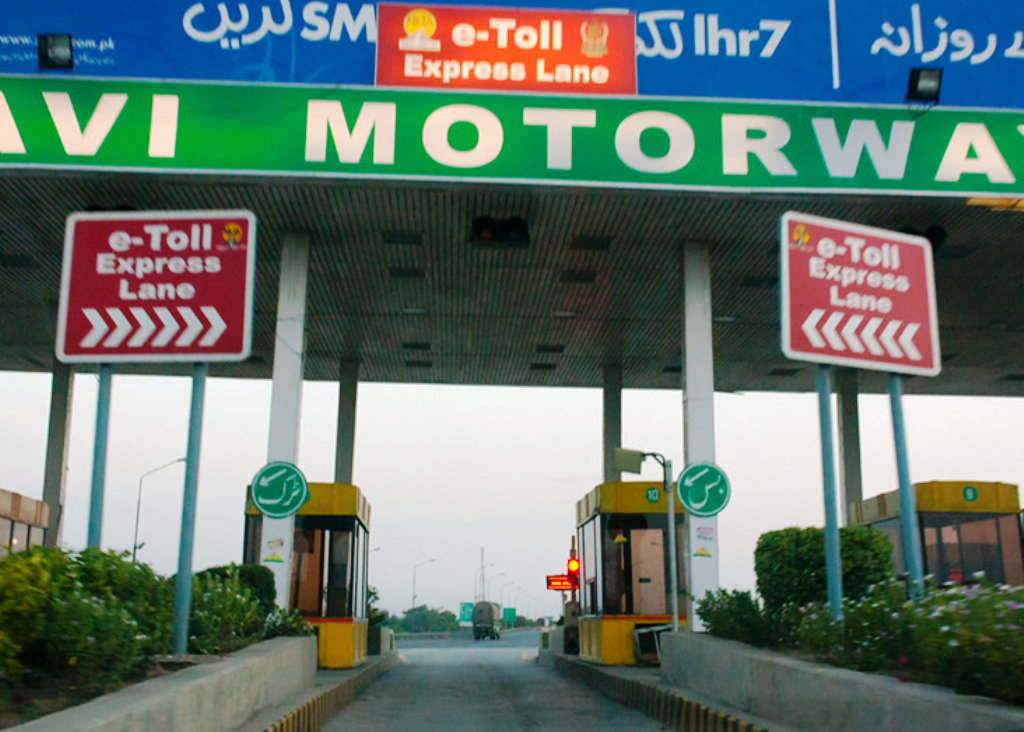}
									
							\end{subfigure}\\
							\begin{subfigure}[t]{0.4\textwidth}
									\centering
									\includegraphics[width=1\textwidth]{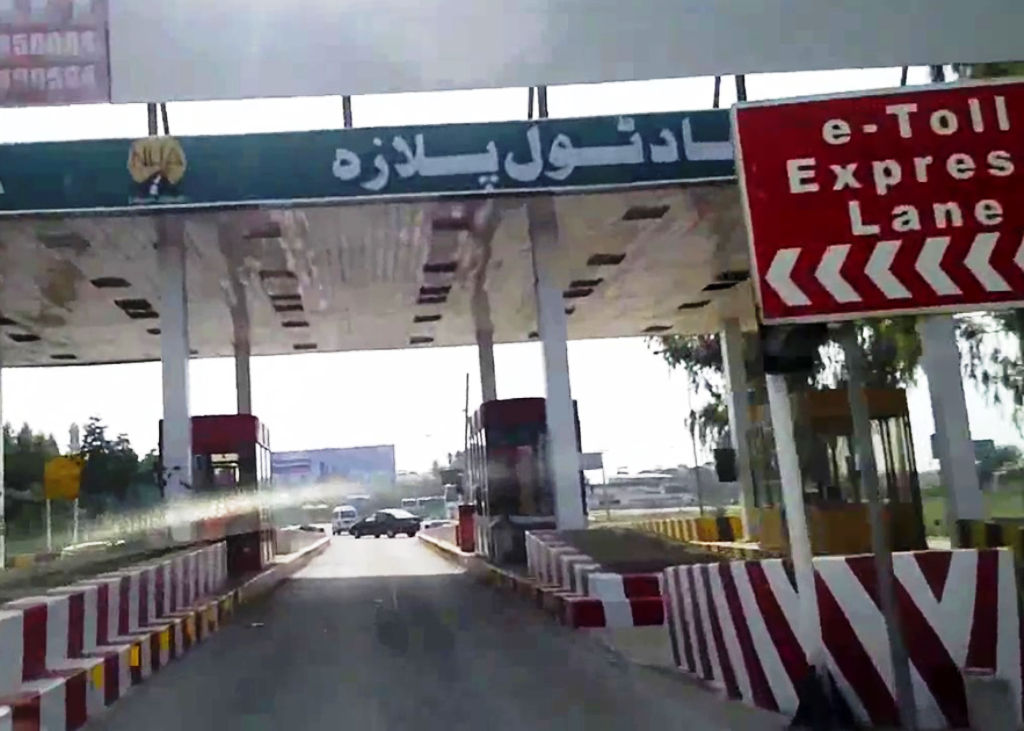}
									
							\end{subfigure}\\
							
					\end{tabular}

    \end{tabular}
    }
		\caption{Toll plaza traffic congestion[left column] and e-tag based express lane [right column]. These scenarios show the importance of automatic license plate recognition.  }
		\label{fig:1}
\end{figure}

\IEEEPARstart{T}{oll} 
tax is one of the important means of revenue generation for the departments responsible for the operation and maintenance of highways and motorways. 
The process of toll tax collection is automated via sensors and cameras in many countries~\cite{kapsch} where the license plates and vehicle types are automatically detected to calculate and deduct the toll tax for vehicles. Among others, the main advantage of such automation is to avoid long queues on toll collection stations or toll plazas of highways on occasions such as weekends and national holidays, as shown in Figure~\ref{fig:1}. 
However, such systems rely on prior knowledge of the specific format of license plates and their prescribed positions on the front of the vehicles. Due to this reason, their performance is most likely to degrade in places where arbitrary formats and positions of license plates are used by most vehicles. 
An existing solution to this problem is the usage of RFID-based e-tag fitted on vehicles. Figure~\ref{fig:1} shows specific lanes on toll plazas where scanners are installed to automatically scan and read the e-tags for tax deduction without stopping the vehicles. 

\begin{figure*}[t!]
\begin{center}

\begin{tabular}{p{2.5cm}p{2.5cm}p{2.5cm}p{2.5cm}p{2.5cm}p{2.5cm}}

\multicolumn{1}{c}{Intra-Class}&
\multicolumn{1}{c}{Non-Uniform}&
\multicolumn{1}{c}{Multiple}&
\multicolumn{1}{c}{Handwritten}&							\multicolumn{1}{c}{Occluded}&
\multicolumn{1}{c}{Damaged}\\

\multicolumn{1}{c}{variations}&
\multicolumn{1}{c}{positions}&
\multicolumn{1}{c}{license plates}&
\multicolumn{1}{c}{license plates}&							
\multicolumn{1}{c}{license plates}&
\multicolumn{1}{c}{license plates}\\

\includegraphics[width=.15\textwidth]{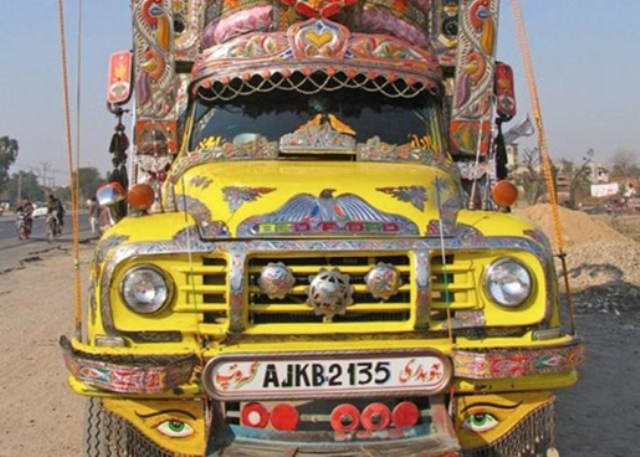}&
\includegraphics[width=.15\textwidth]{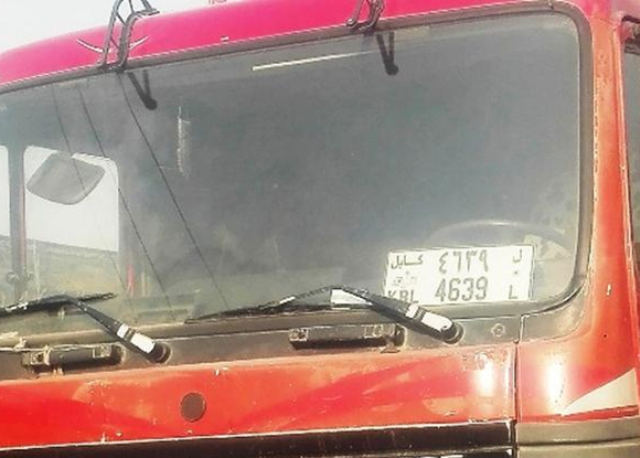}&
\includegraphics[width=.15\textwidth]{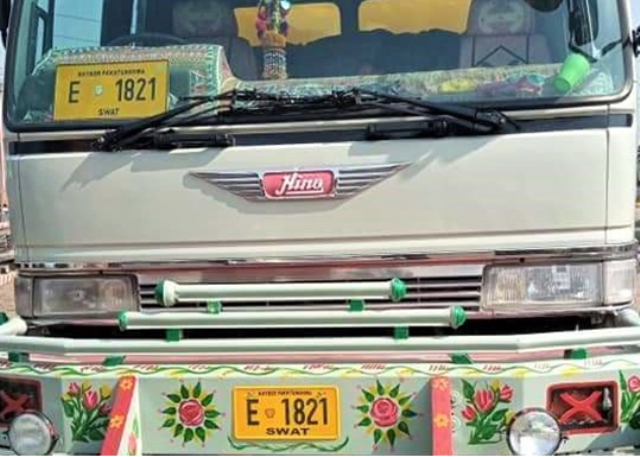}&
\includegraphics[width=.15\textwidth]{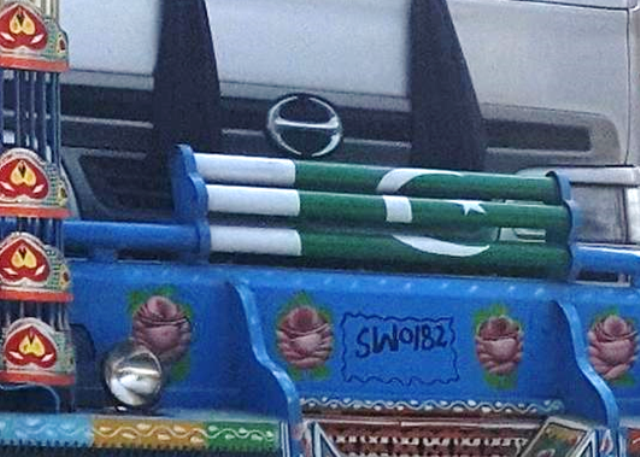}&
\includegraphics[width=.15\textwidth]{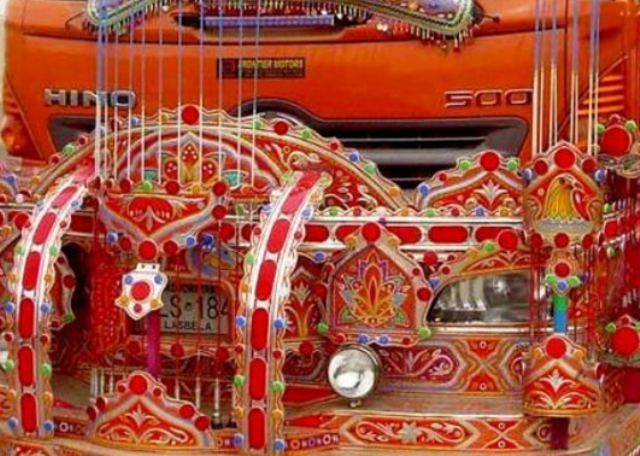}&
\includegraphics[width=.15\textwidth]{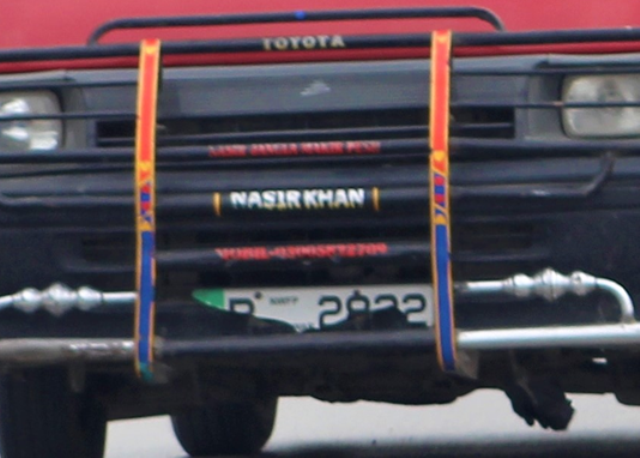}\\

\includegraphics[width=.15\textwidth]{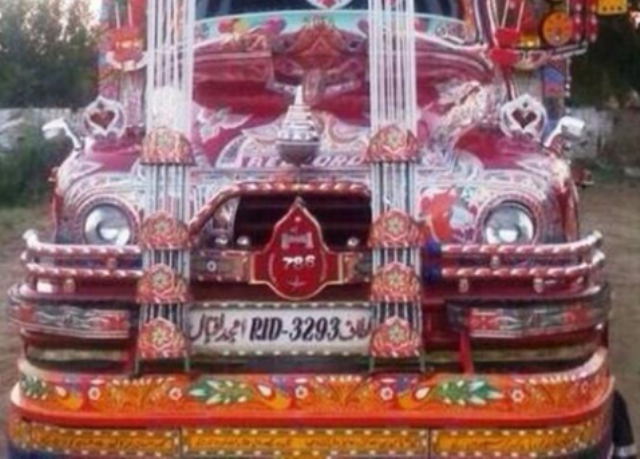}&
\includegraphics[width=.15\textwidth]{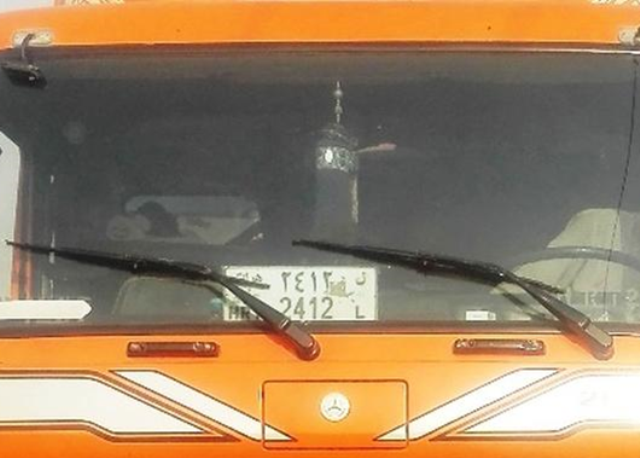}&
\includegraphics[width=.15\textwidth]{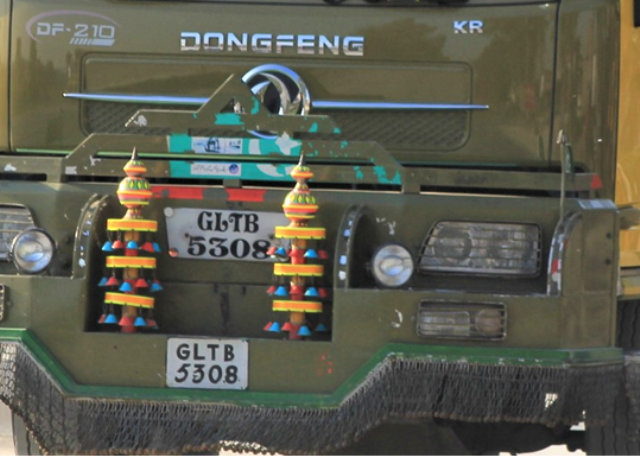}&
\includegraphics[width=.15\textwidth]{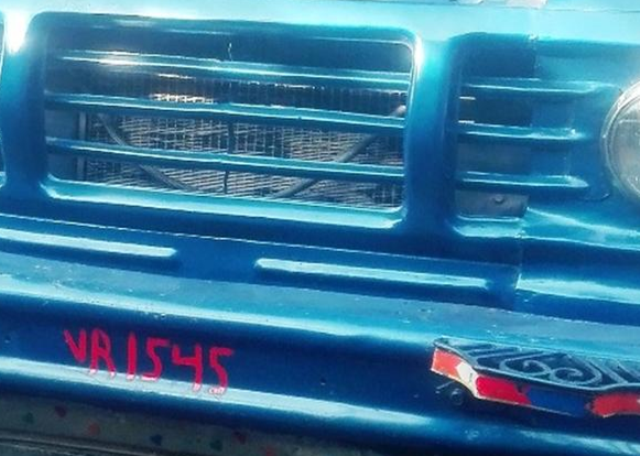}&
\includegraphics[width=.15\textwidth]{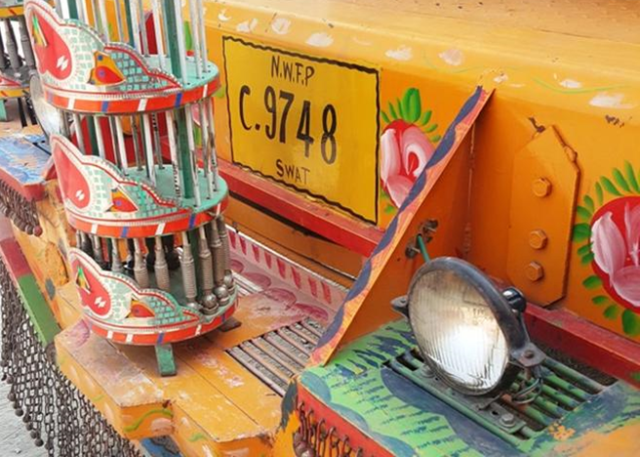}&
\includegraphics[width=.15\textwidth]{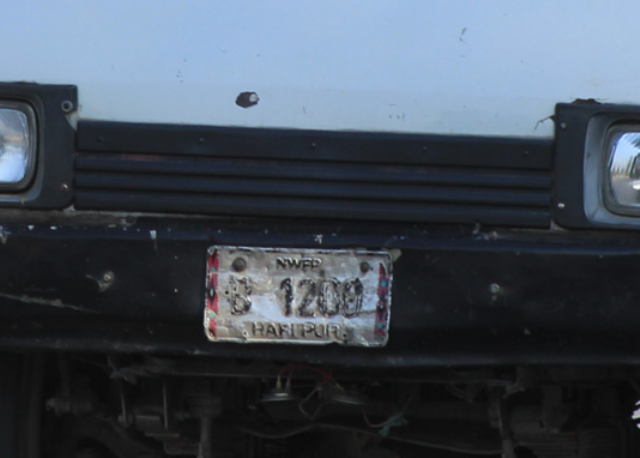}\\

\includegraphics[width=.15\textwidth]{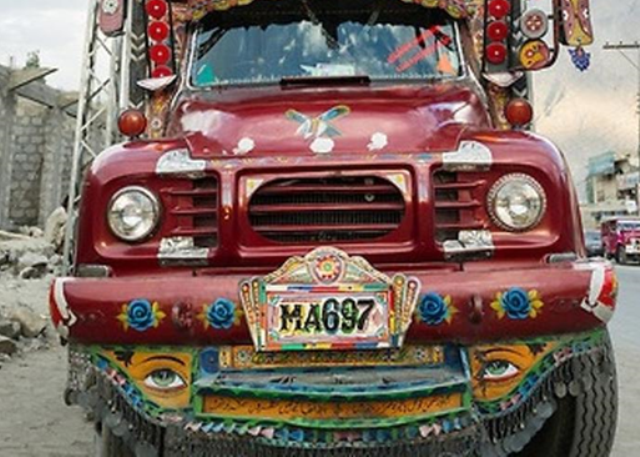}&
\includegraphics[width=.15\textwidth]{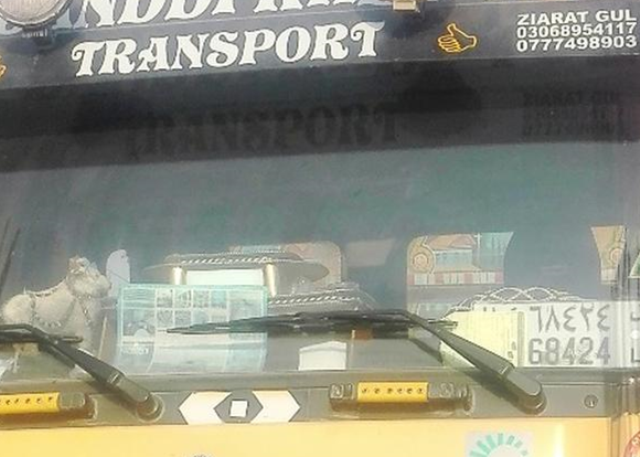}&
\includegraphics[width=.15\textwidth]{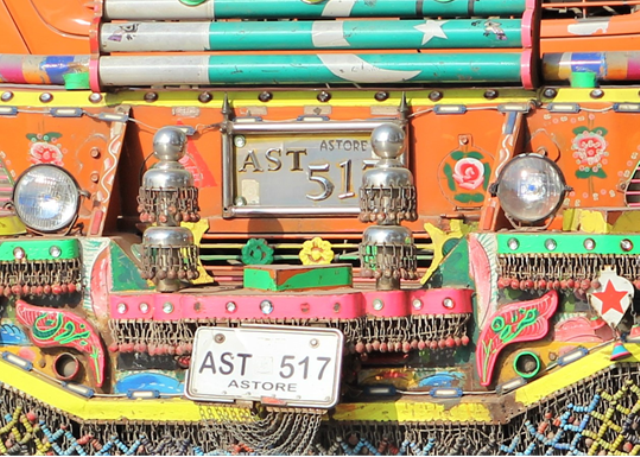}&
\includegraphics[width=.15\textwidth]{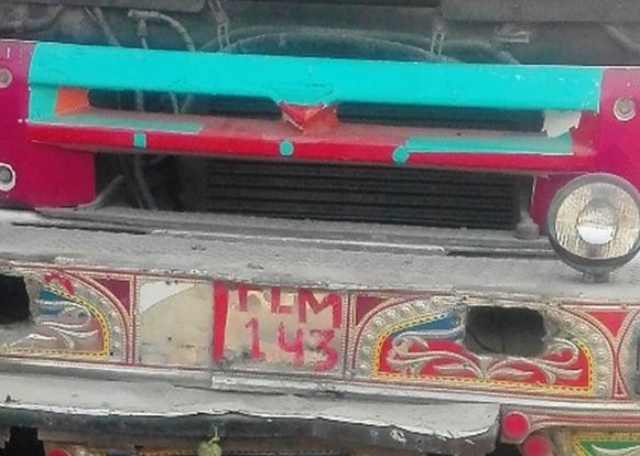}&
\includegraphics[width=.15\textwidth]{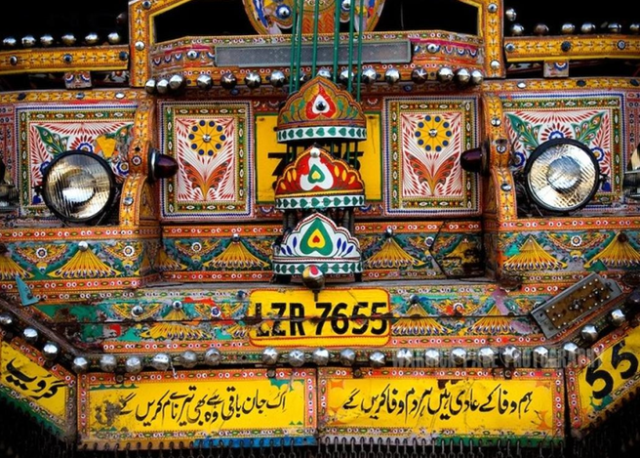}&
\includegraphics[width=.15\textwidth]{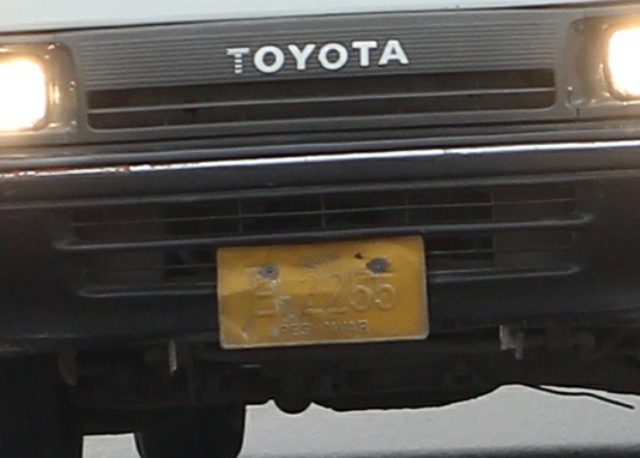}\\

\includegraphics[width=.15\textwidth]{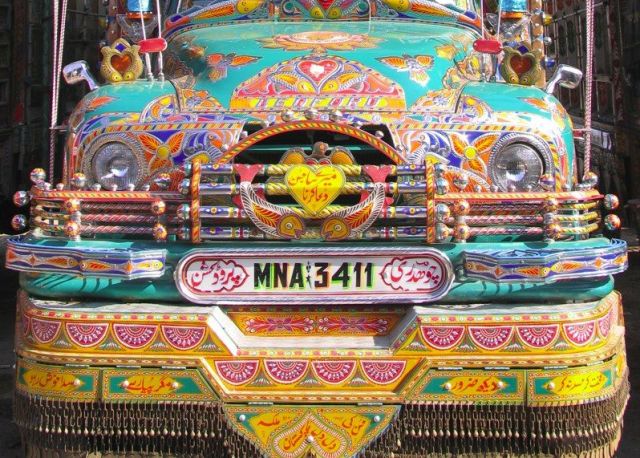}&
\includegraphics[width=.15\textwidth]{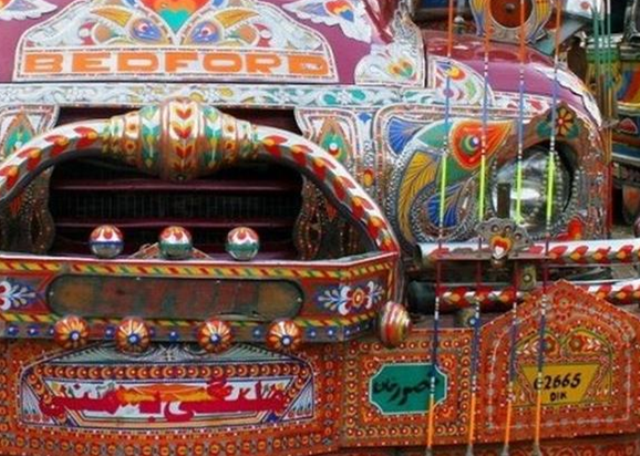}&
\includegraphics[width=.15\textwidth]{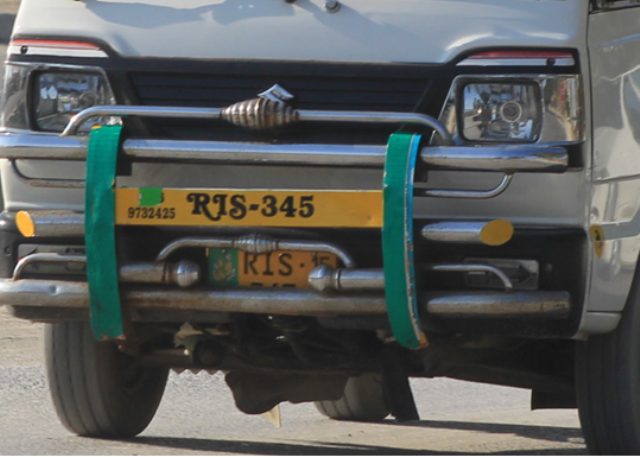}&
\includegraphics[width=.15\textwidth]{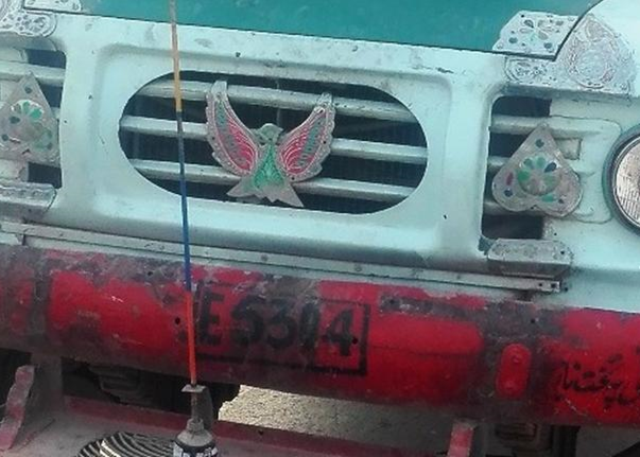}&
\includegraphics[width=.15\textwidth]{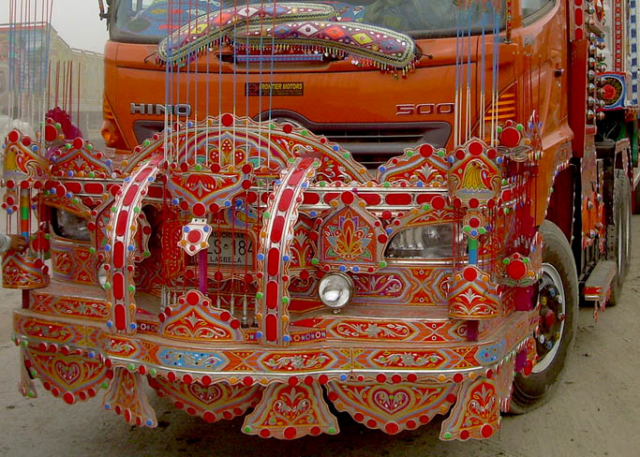}&
\includegraphics[width=.15\textwidth]{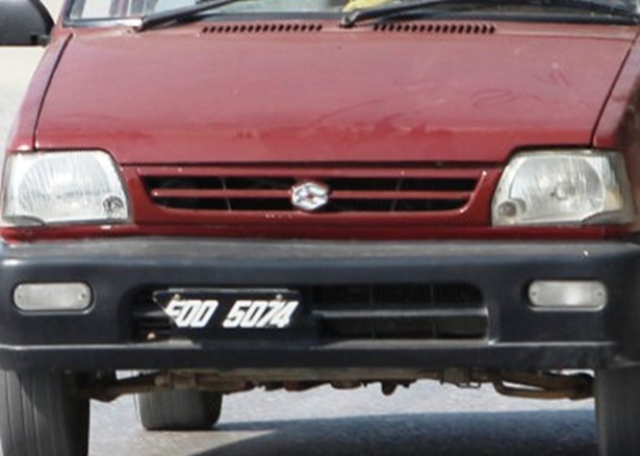}\\
								
\end{tabular}
\end{center}
\caption{Exemplar images of various challenges caused by intra-class type variations due to decorations, license plate positions are not coherent, more than one license plate displayed at vehicle's front, license plates are occluded by decorations and broken or faded license plates.
}
\label{fig:2}
\end{figure*}

However, this comes with an additional cost of installing extra infrastructure on toll plazas and e-tags on the vehicles. In addition, e-tags are only installed on regular users of motorways that constitute a minor portion of the whole traffic volume. 

Leveraging upon the installed surveillance cameras of toll plazas, we propose an image-based solution for toll tax collection. Such a solution is motivated by the recent state-of-the-art results achieved by the usage of convolution neural network (CNN) for image-based vehicle type\cite{chaitanya2021object} and license plate recognition~\cite{wang2020rethinking}. Our framework calculates the toll tax of a vehicle from its image using three steps strategy. The first step deals with the localization and type recognition of the vehicle in the image. The region of the image that depicts the vehicle is then searched for the license plate. The final step is then performed on the localized license plate by reading its characters and digits. All three steps are performed with a CNN-based object localization framework.  
However, the task of the image-based vehicle and license plate recognition becomes challenging due to variations found in the images of the Pakistani vehicles. Following is the list of challenges whose exemplar images are also shown in Figure~\ref{fig:2}. 
\begin{itemize}
    \item \textbf{Intra-class variations}: Our framework aims at the vehicle type recognition from its front that is affected by the traditional decorations~\cite{truckart} on the vehicles. Such decorations cause dissimilarities among vehicles of the type.
    \item \textbf{Background clutter}: The decorations surrounding the license plate cause background clutter, thus making it difficult to get detected.
    \item \textbf{Non-uniform position}: In the vast majority of the vehicles, the official license plate positioning is not followed, thus making it difficult to detect the license plate. 
    \item \textbf{Multiple license plates}: It is also noted that in some vehicles, two license plates are displayed where one is on the front side, and the other is displayed inside the windshield.
    \item \textbf{Non-uniform fonts}: The variations in fonts caused by the use of non-official license plates make it difficult to recognize the characters and digits on the license plates.
    \item \textbf{Partial occlusions}: Due to the dusty environment, the license plates are always partially occluded with dust and mud.
    \item \textbf{Damaged license plates}: In some cases, the non-official license plates are damaged, which causes the missing of some digits or characters.
    \item \textbf{Lack of license plates}: The license plate can even be missing in cases like new vehicles where at the place of the license plate, a message of \enquote{Applied for} is displayed.
\end{itemize}

Due to these challenges, our proposed framework consists of three individual steps, which are (i) vehicle detection and type recognition, (ii) license plate detection, and (iii) license plate reading. We treat each step as an object detection and recognition problem, where we train a separate model for each of them. Consequently, all three models are applied in a cascaded manner during the run time to recognize both the vehicle and its license plate for calculating the toll tax. In this regard, the following are our novel/significant contributions.
\begin{enumerate}
    \item We collect a dataset of vehicles that consists of 10K images. This is one of the largest and most diverse image datasets of vehicles to the best of our knowledge.
    \item From these images, we generate three sub-datasets each for vehicles type recognition, license plate detection, digits, and character recognition of license plates.
    \item We extensively evaluate the performance of several object detection frameworks, which are YOLOv4 and Tiny YOLOv4, YOLOv3 and Tiny YOLOv3, YOLOv2, and FasterRCNN for each of the three steps.
    \item For IoT-based and real-time solutions, we also deploy our proposed framework on a Raspberry Pi along with a Pi camera and a graphical user interface (GUI). 
\end{enumerate}
The rest of the paper outlines literature review in Section~\ref{sec:literature}, dataset in Section~\ref{sec:dataset}, the proposed methodology in Section~\ref{sec:method}, experimental results in Section~\ref{sec:results} and conclusion in Section~\ref{sec:conclusion}.

\section{Literature Review}
\label{sec:literature}
Since we propose to perform both vehicle and number plate recognition, this section summarizes some literature on both problems. 

\subsection{Vehicle type recognition}
Vehicle type recognition deals with the coarse-grained classification of a vehicle where the aim is to put it into one of the classes such as car, bus, van, and truck, \etc The image-based vehicle type recognition methods are either based on the so-called handcrafted features or use convolutional neural networks. We give a summary of some of the techniques from both groups.

\textbf{Handcrafted features}: 
The Scale-invariant Features Transform (SIFT)~\cite{lowe2004distinctive} is used by~\cite{6738890} for appearance representation while the relative positions of SIFT feature as a structural feature to deal with the variations found in vehicle images. These features are then utilized via multiple kernel learning for vehicle type recognition. A  discriminative shape descriptor is developed with the help of a modified SIFT and edge points for vehicle type recognition~\cite{ma2005edge}. In a similar manner, vehicle type recognition is performed using SIFT by other methods~\cite{dlagnekov2005recognizing},~\cite{conos2007recognition}, ~\cite{psyllos2011vehicle}. 
Sun~\etal~\cite{sun2017vehicle} employ a two-step classification strategy where in the first step, the detected vehicle is classified as a small or large vehicle by using global features generated with improved Canny Edge detection. In the second step, the finer type is predicted for the vehicle via Gabor wavelet kernels extracted at five scales with eight orientations.

Peng~\etal~\cite{6266414} utilize the location of the license plate as a prior to extracting the front of the vehicle. The feature vector of the extracted vehicle front is then composed with the help of Eigenvectors. K-means clustering is then applied over these vectors in order to represent a vehicle type with a cluster center. A given test sample is then classified by calculating the distance of its feature vector to that of a cluster center. A follow-up work~\cite{peng2013vehicle} improves this method by enriching it with other information queues such as type-specific license plate color and background subtraction. 
Others handcrafted features-based for vehicle type recognition use gradient and edges~\cite{petrovic2004analysis}, edge direction-based deformable templates~\cite{jolly1996vehicle}, and Gaussian Mixture Model~\cite{kuwabara2008vehicle}.

\textbf{CNN-based methods}:
A two-steps strategy is proposed~\cite{doi:10.3141/2645-13} for vehicle type recognition, where in the first step, class proposal regions are generated for vehicle localization. In the second step, descriptors are obtained for these regions using CNN embeddings, where the vehicle type recognition is then performed using an SVM over these embeddings. The vehicles types used in the study include cars, SUVs, single and double trailer trucks, where for cars and SUVs, the achieved precision is 95\%, and for the trucks, it ranges between 92\% and 94\%.
Vehicle type recognition is performed~\cite{wang2017vehicle} in surveillance videos using CNN. However, the training image data comes from the web, due to which the authors modify the objective function to get a generalized transfer learning for vehicle type recognition. In order to reduce the cost of manual annotation of the surveillance videos, A performance evaluation of the traditional method of SVM on SIFT features against a deep neural network (DNN) is done by~\cite{huttunen2016car} DNN performs the best. 

Kim \& Lim~\cite{kim2017vehicle} train various CNN models on randomly sampled images from the training set. The number of sampled images is equal to half the size of the training set.
The so-called active learning technique is used by~\cite{8961120} where the network training is started with relatively smaller and labeled training data. Further data is then automatically selected by the network, both with high and low entropy for the retraining process. However, in training, data augmentation is also used to generate the augmented images by flipping and rotating the original photos. Each trained model generates a prediction for a test image, where a weighted voting process calculates the final prediction. A similar strategy~\cite{taek2017deep} is adopted, where the ensemble of the so-called local and global networks is used such that the local networks are trained on subsets of training set while the global is trained on the entire training dataset. The inference for a test image where a single prediction is selected from local networks. The final outcome is then made by aggregating the predictions of all global networks and the ones chosen from the local networks.

Since, at toll tax collection, the vehicles line up properly at the toll plazas, we aim to recognize a vehicle from the front. Nonetheless, our approach to performing image-based vehicle type recognition is different from all the previously stated approaches in the following manner.
\begin{itemize}
    \item We face the problem of severe clutter caused by the traditional decoration measures.
    \item Such decorative measures also cause high intra-class variations, such that vehicles belonging to the same class look very different from one another.
    \item Finally, the high dusty conditions at highways in Pakistan also cause a major challenge for vehicle type recognition
\end{itemize}

\subsection{License plate detection and recognition}
The second important step of our proposed pipeline is the localization of the license plate and then recognizing the digits and characters. From a detected vehicle in the image, the aim is to detect, localize, extract and then read its license plate. Vehicle license plate detection and reading is a well-researched problem; however, it is still an active area of research. We give an overview of both the handcrafted features-based and CNN-based methods in the following. 

\textbf{Handcrafted features}:
Farmanullah~\etal~\cite{ullah2019barrier} perform detection of the license plates using image processing operators. The character recognition on the detected license plates is then performed by evaluating various machine learning classifiers over the geometric features of the characters. 
The Histogram of Oriented Gradients (HoG)~\cite{dalal2005histograms} with a Support Vector Machine (SVM) in order to detect and recognize license plates in challenging weather conditions and uneven illumination~\cite{rio2019effects}. Other methods of license plate recognition use SIFT~\cite{zahedi2011license}~\cite{wang2015license}, Maximally Stable Extremal Regions (MSER)~\cite{gou2014license}~\cite{li2012vehicle} and a combination of MSER and SIFT~\cite{lim2010detection}. 
\begin{figure*}[t!]
\begin{center}

\begin{tabular}{p{2.5cm}p{2.5cm}p{2.5cm}p{2.5cm}p{2.5cm}p{2.5cm}}

\multicolumn{1}{c}{Buses}&
\multicolumn{1}{c}{Cars}&
\multicolumn{1}{c}{Carry vans}&
\multicolumn{1}{c}{Truck Type 1}&							
\multicolumn{1}{c}{Truck Type 2}&
\multicolumn{1}{c}{Vans}\\

\includegraphics[width=.15\textwidth]{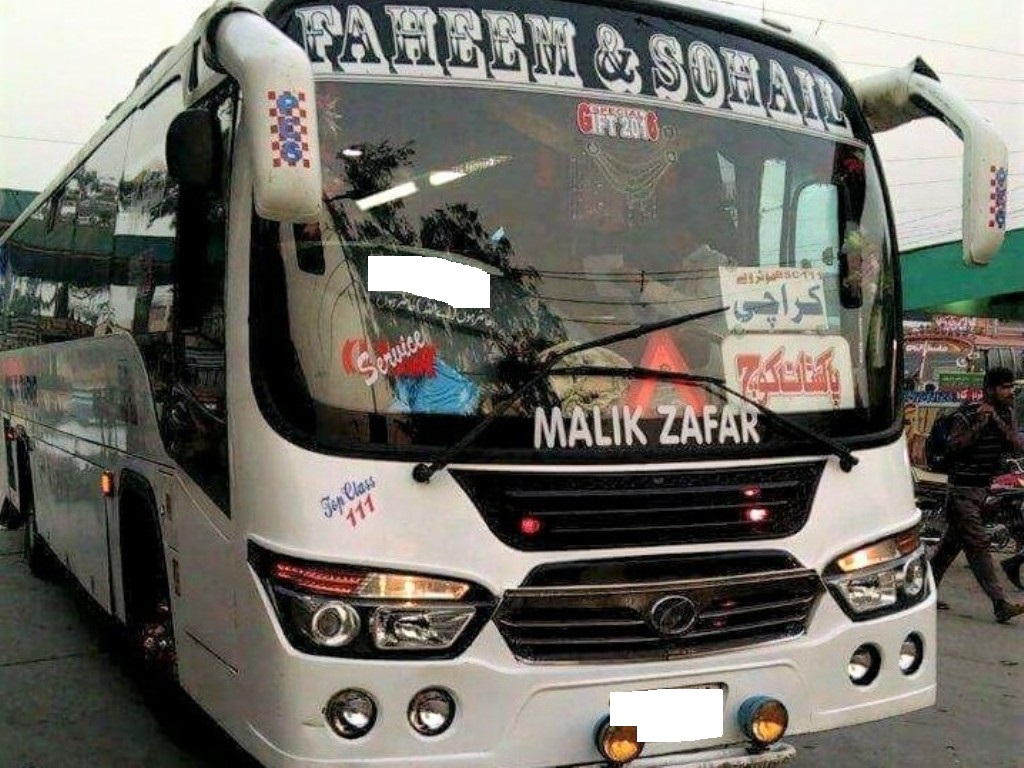}&
\includegraphics[width=.15\textwidth]{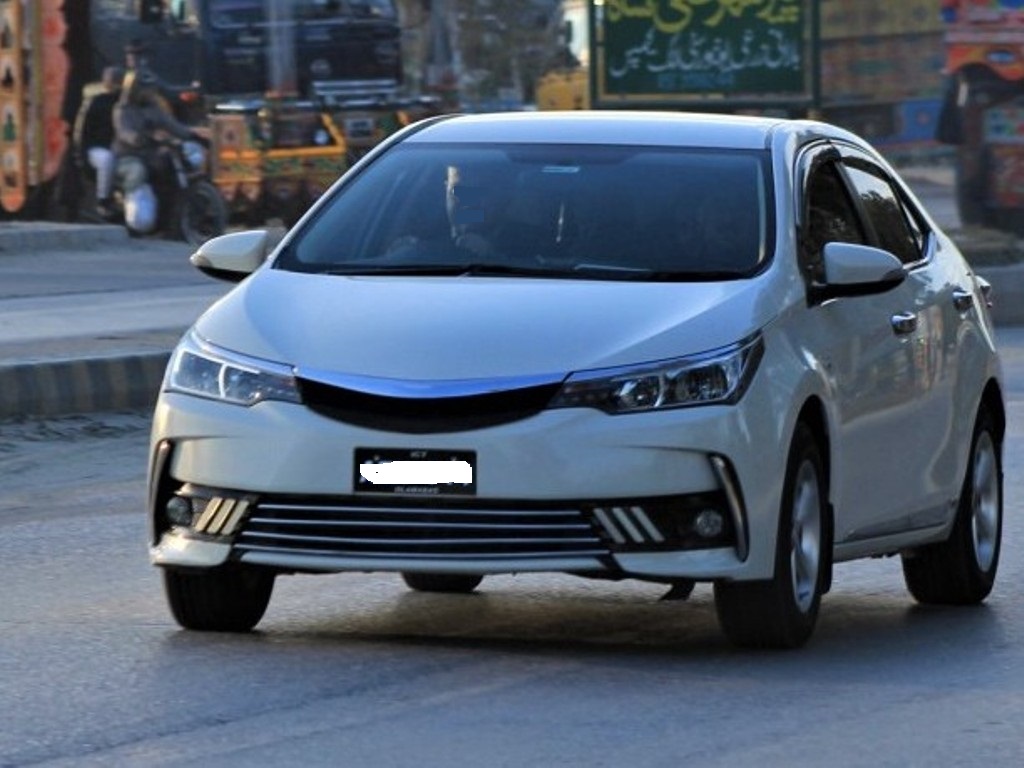}&
\includegraphics[width=.15\textwidth]{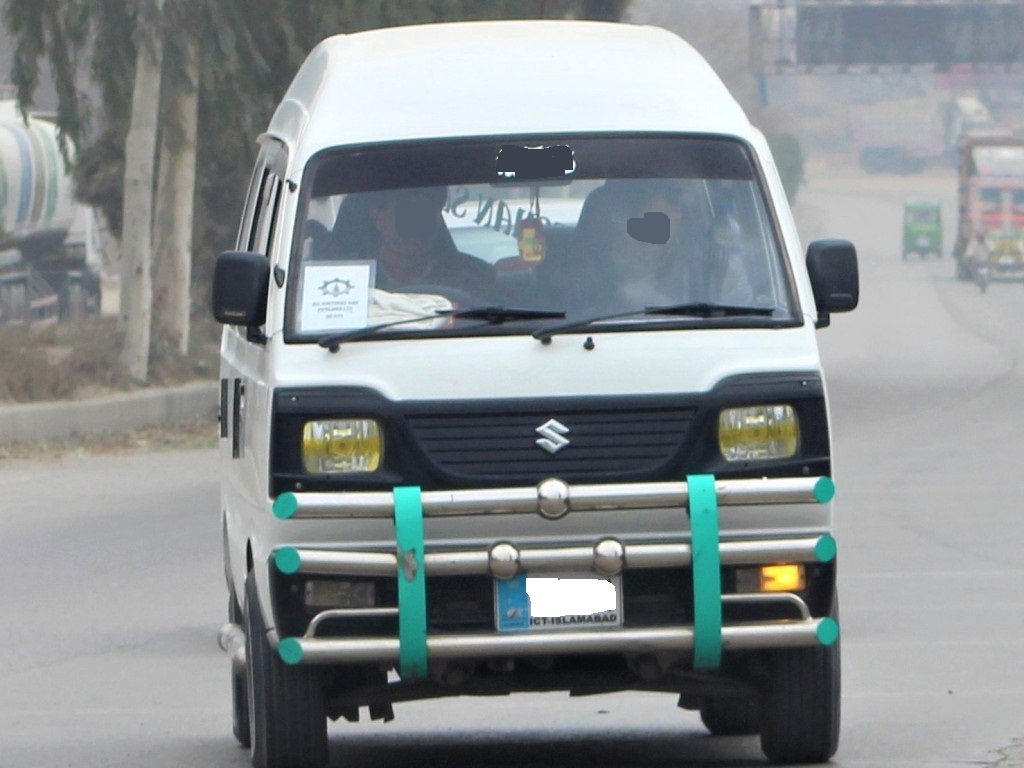}&
\includegraphics[width=.15\textwidth]{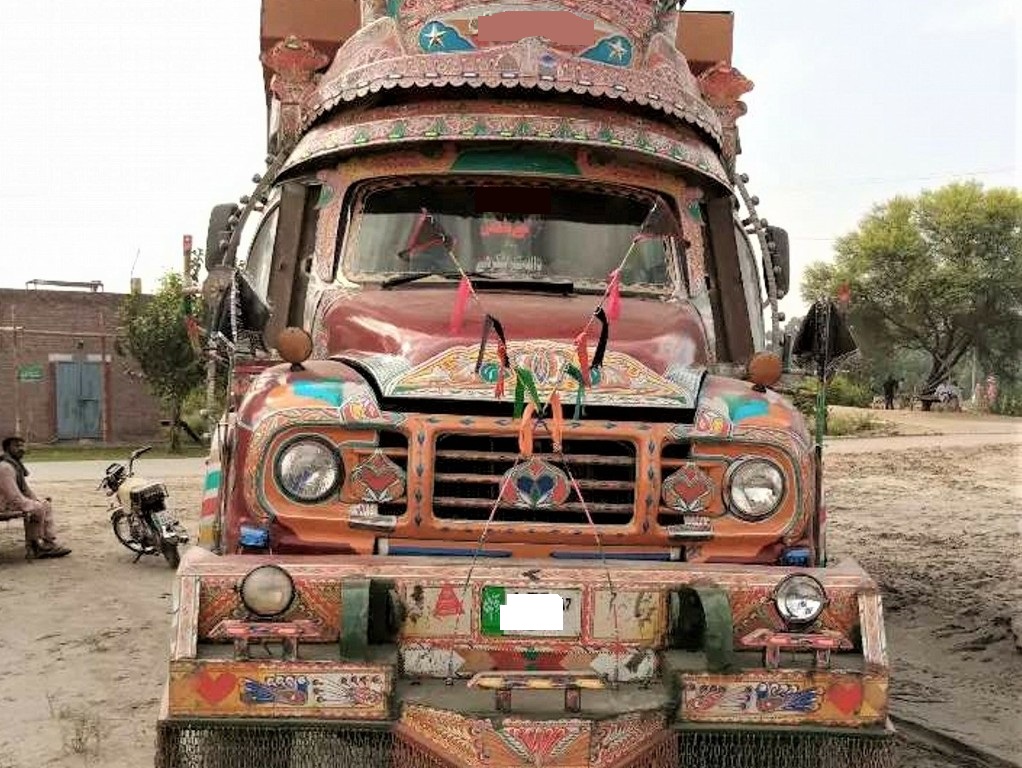}&
\includegraphics[width=.15\textwidth]{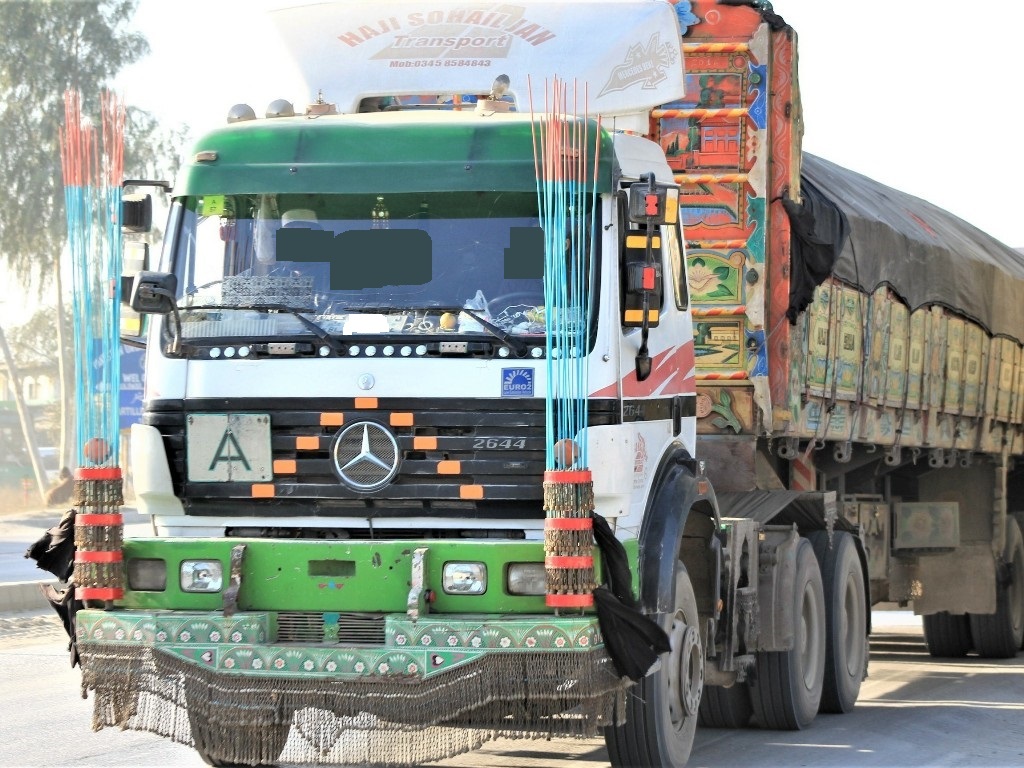}&
\includegraphics[width=.15\textwidth]{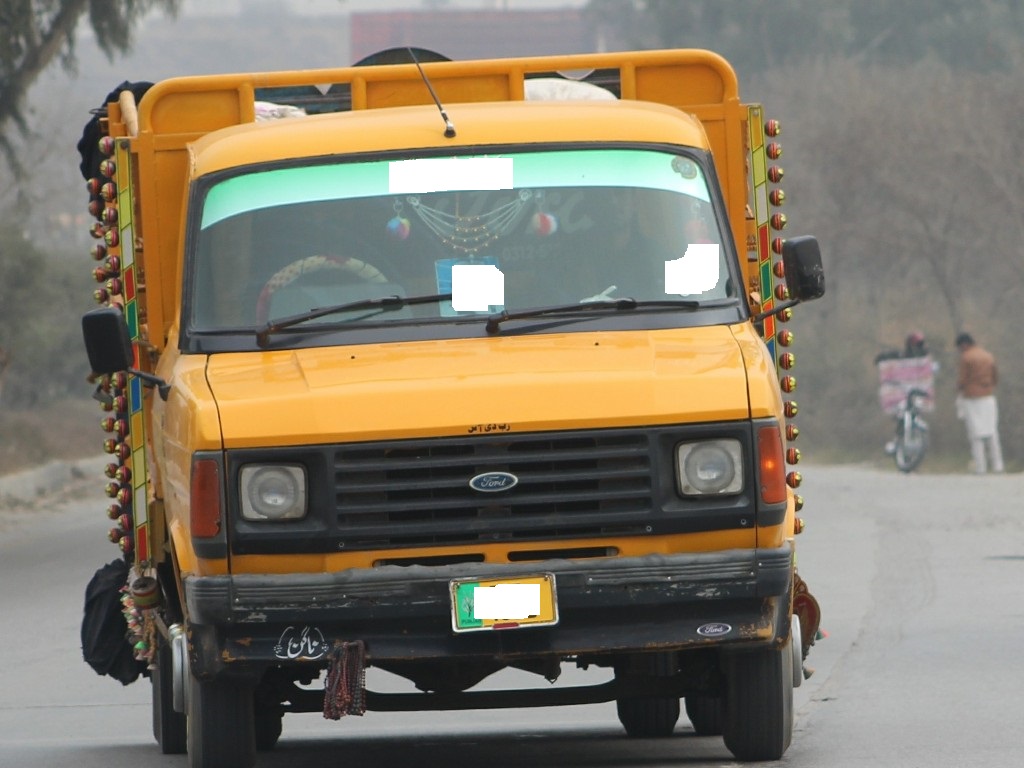}\\

\includegraphics[width=.15\textwidth]{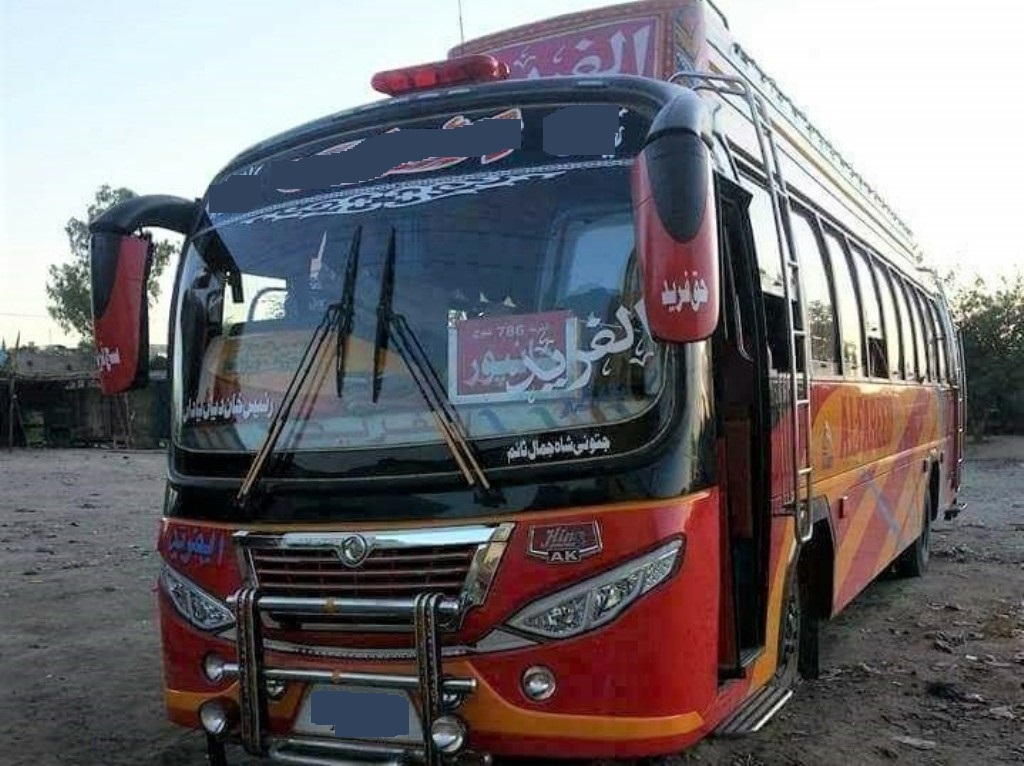}&
\includegraphics[width=.15\textwidth]{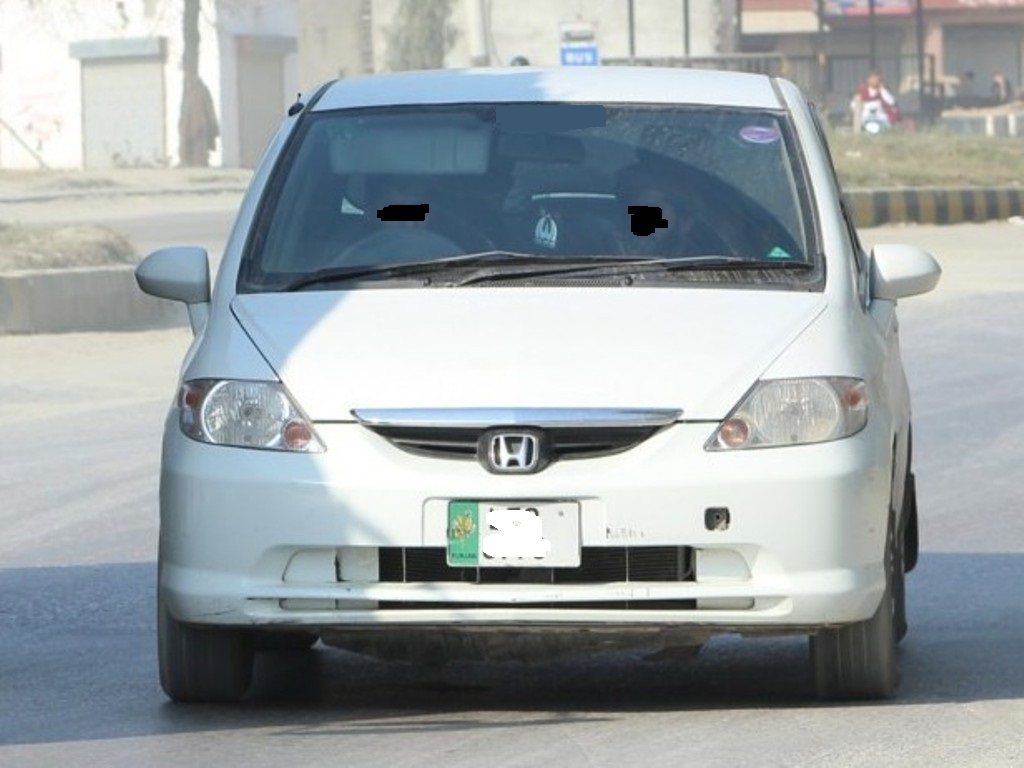}&
\includegraphics[width=.15\textwidth]{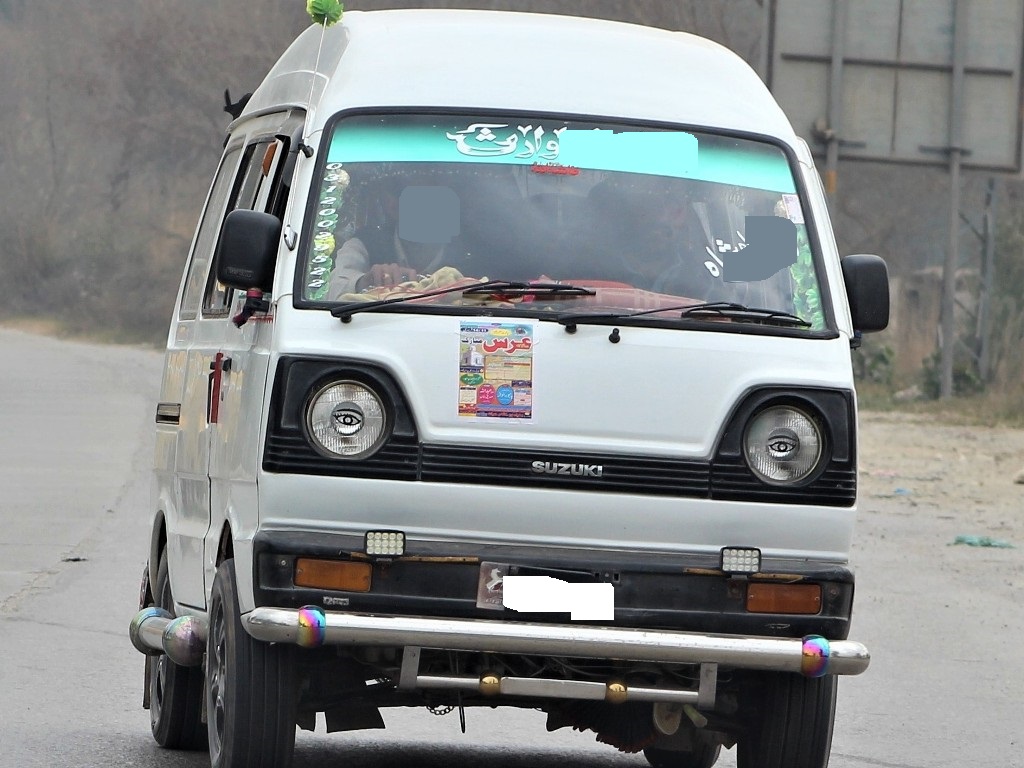}&
\includegraphics[width=.15\textwidth]{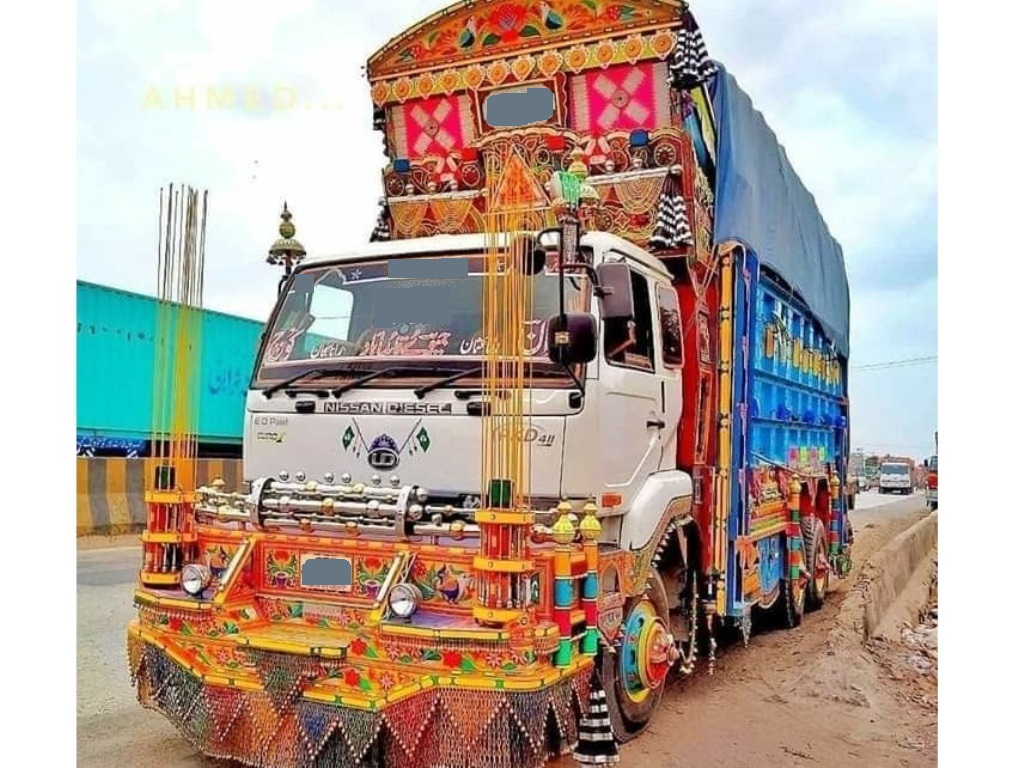}&
\includegraphics[width=.15\textwidth]{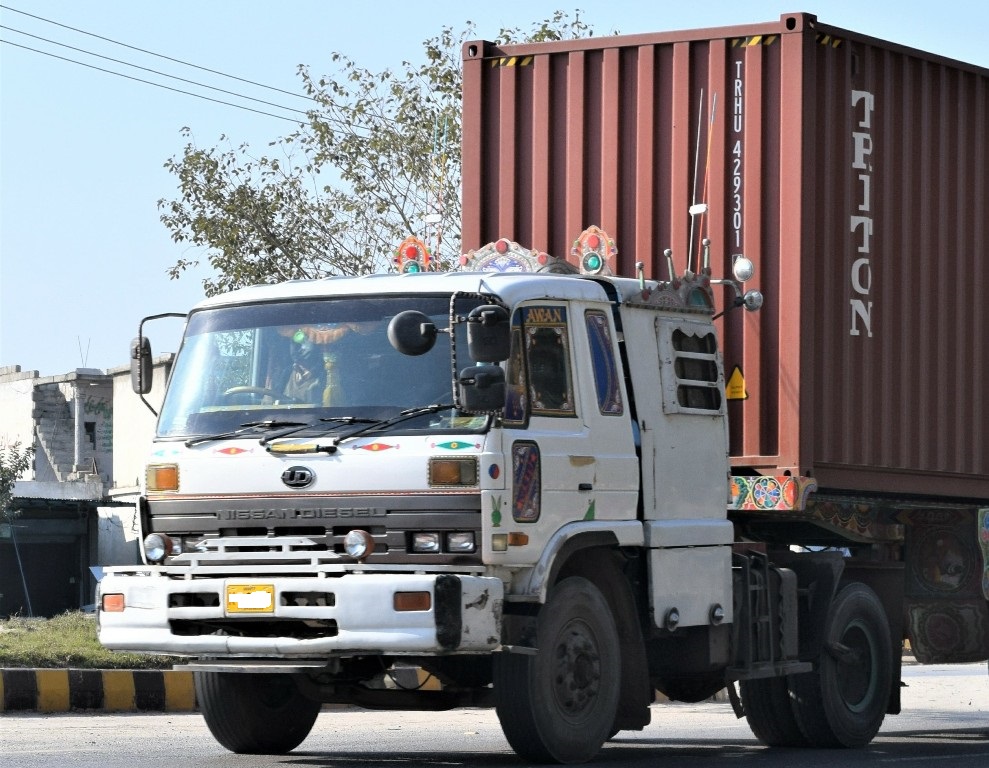}&
\includegraphics[width=.15\textwidth]{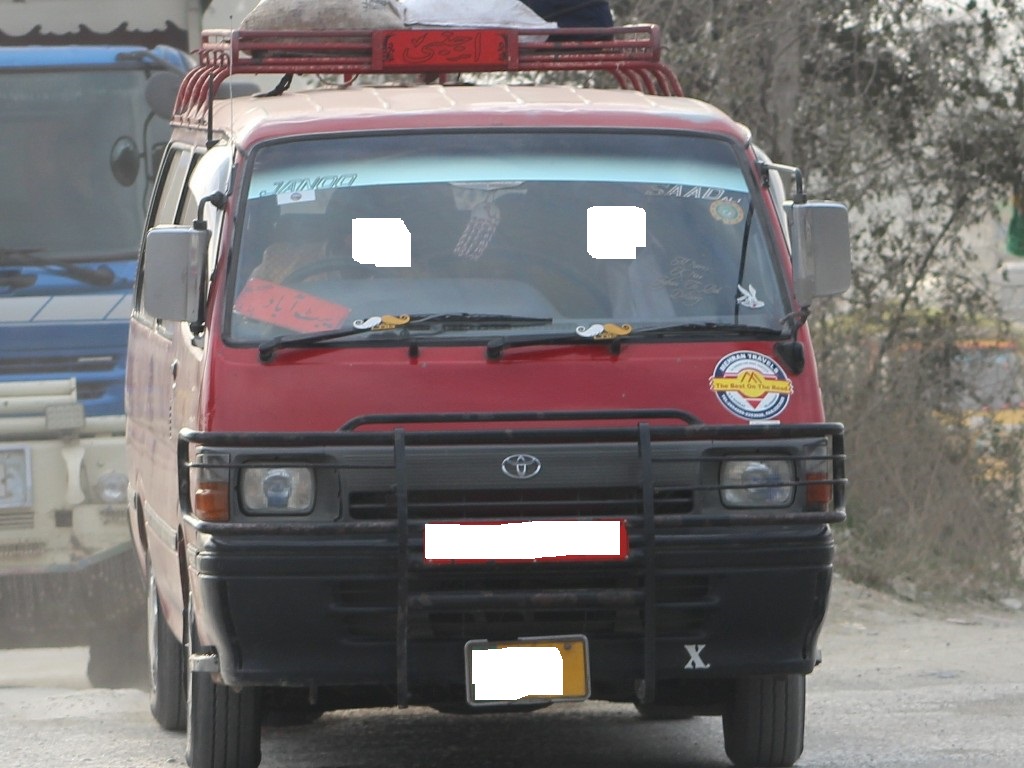}\\

\includegraphics[width=.15\textwidth]{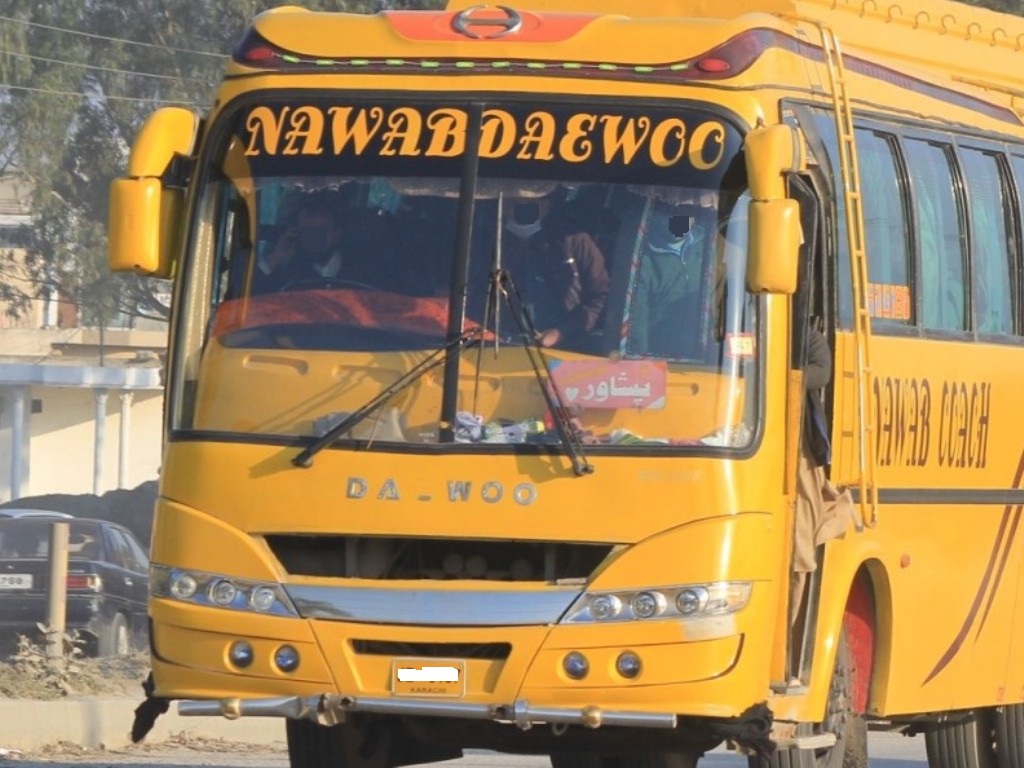}&
\includegraphics[width=.15\textwidth]{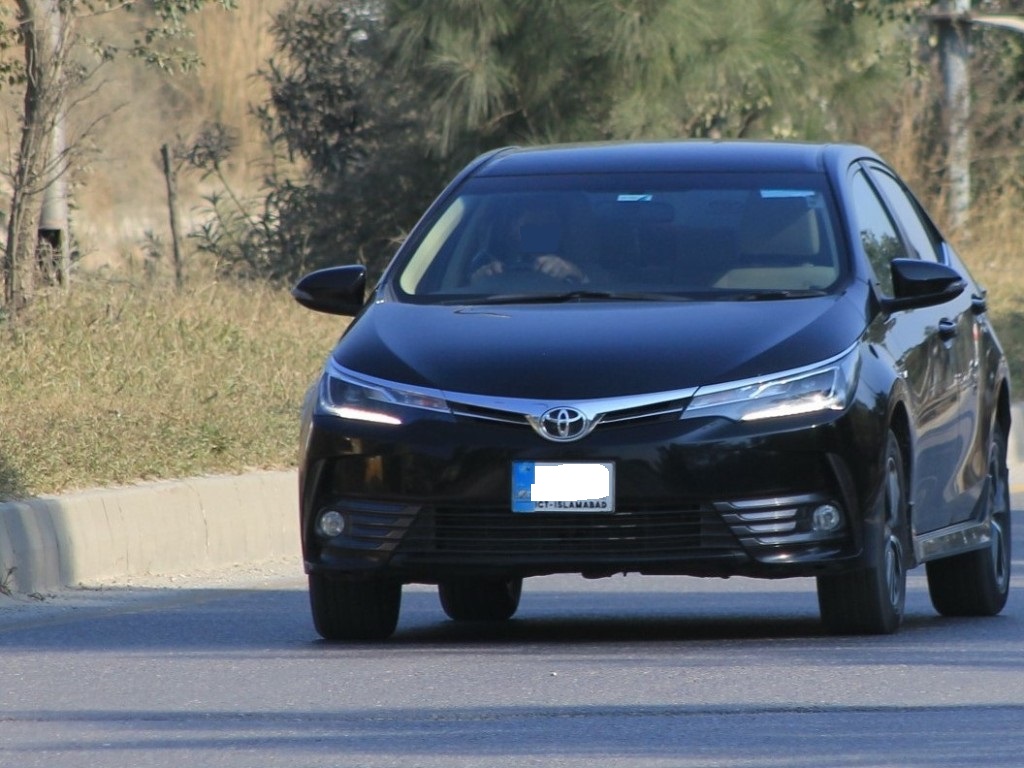}&
\includegraphics[width=.15\textwidth]{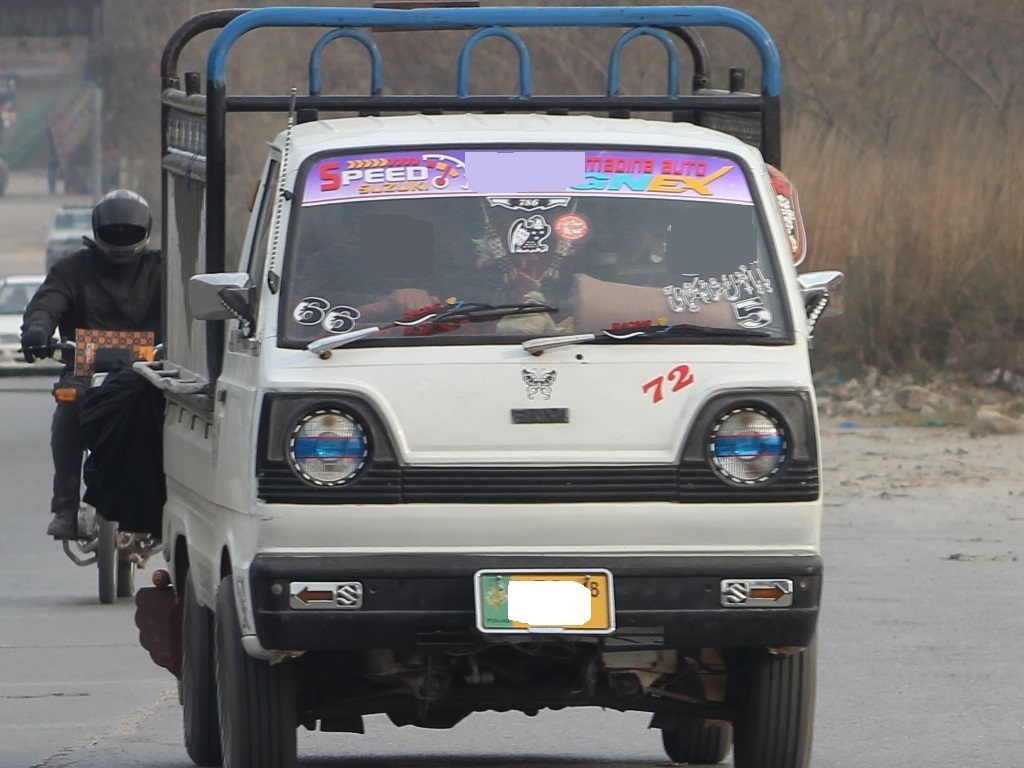}&
\includegraphics[width=.15\textwidth]{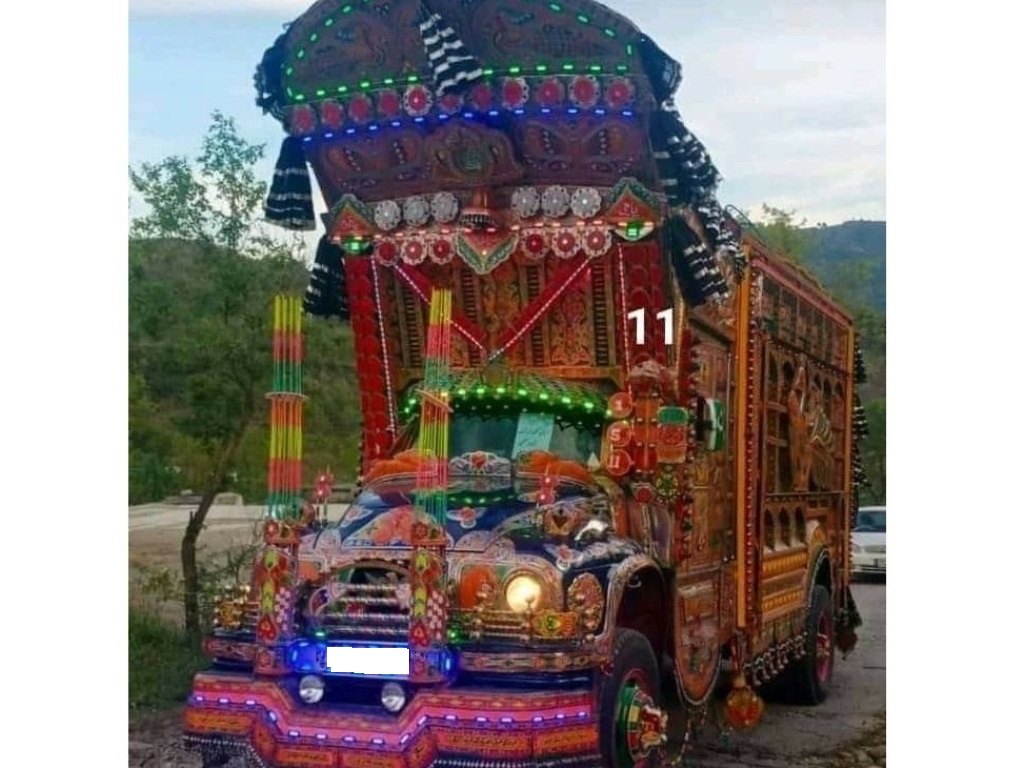}&
\includegraphics[width=.15\textwidth]{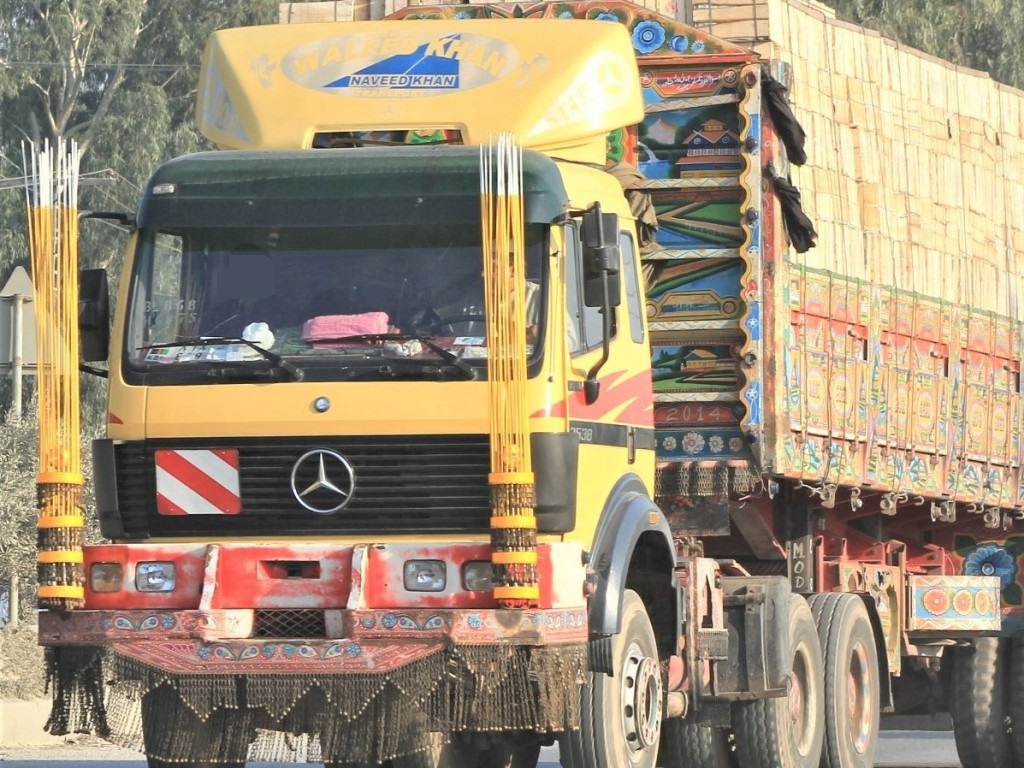}&
\includegraphics[width=.15\textwidth]{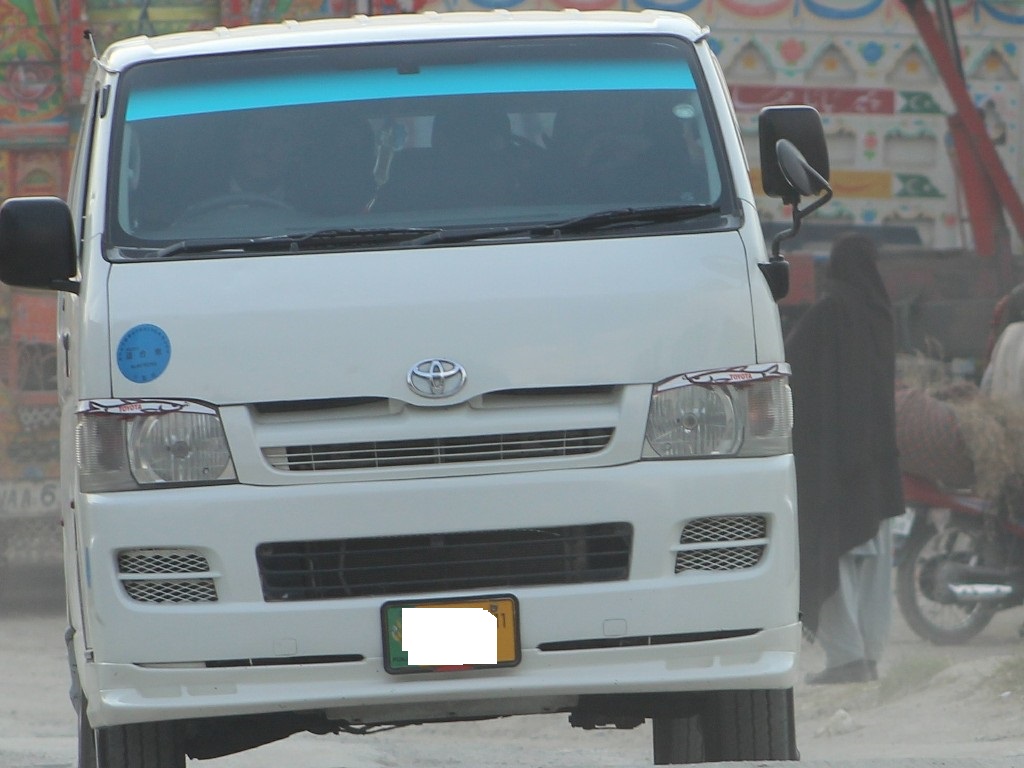}\\

\includegraphics[width=.15\textwidth]{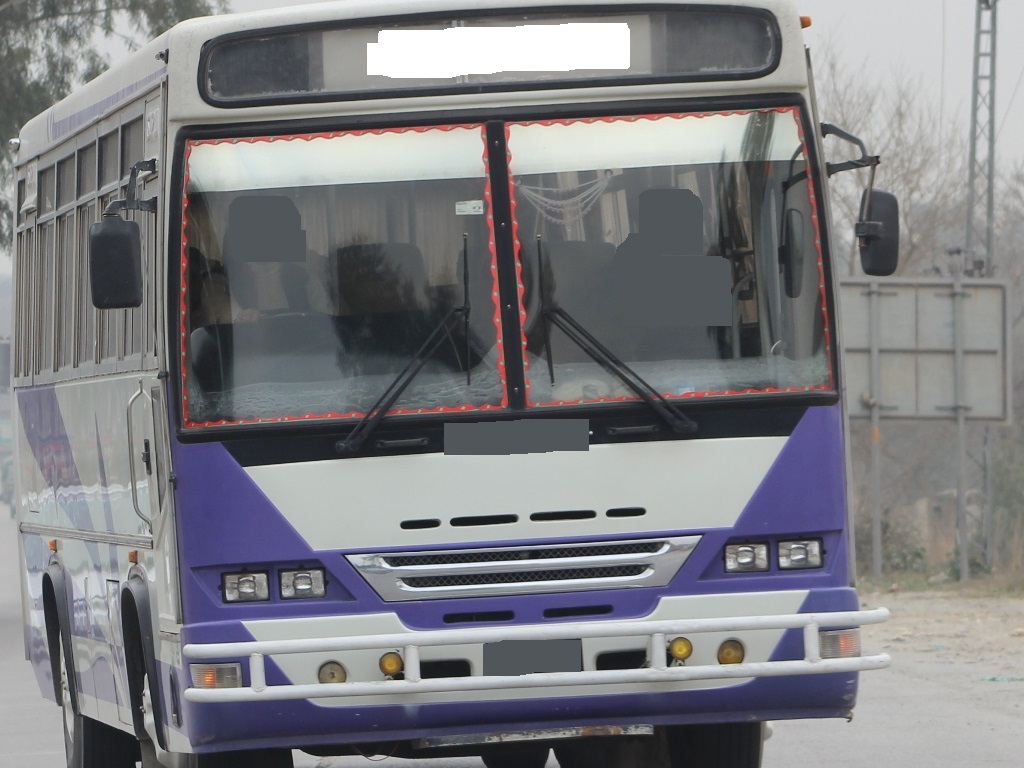}&
\includegraphics[width=.15\textwidth]{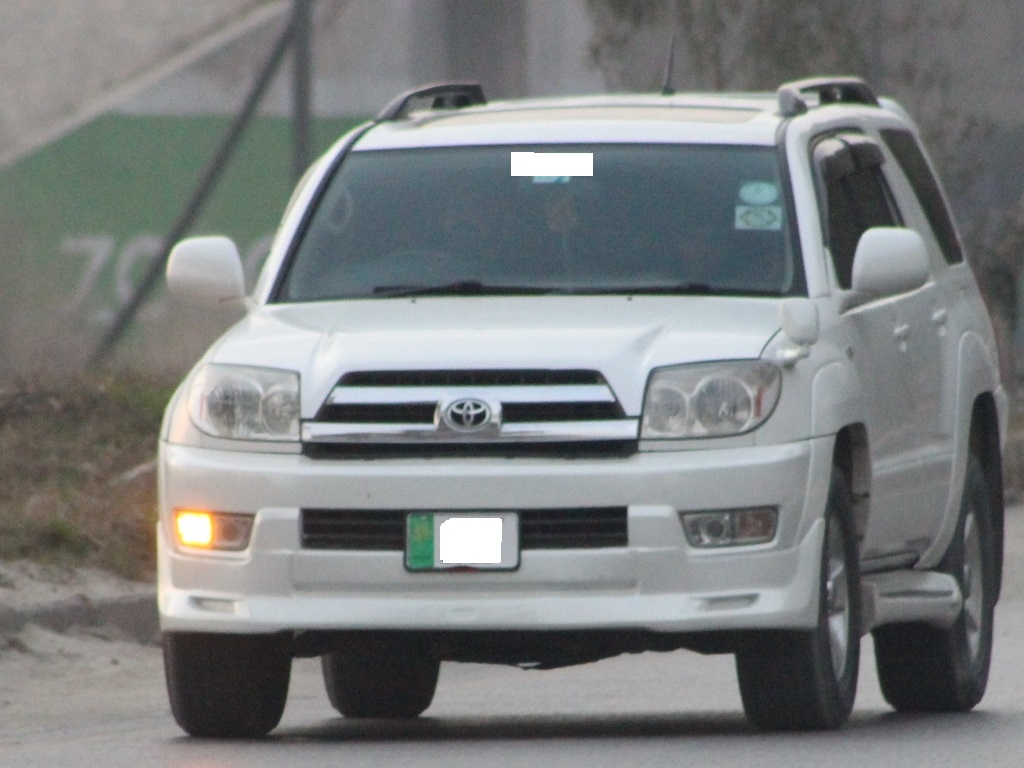}&
\includegraphics[width=.15\textwidth]{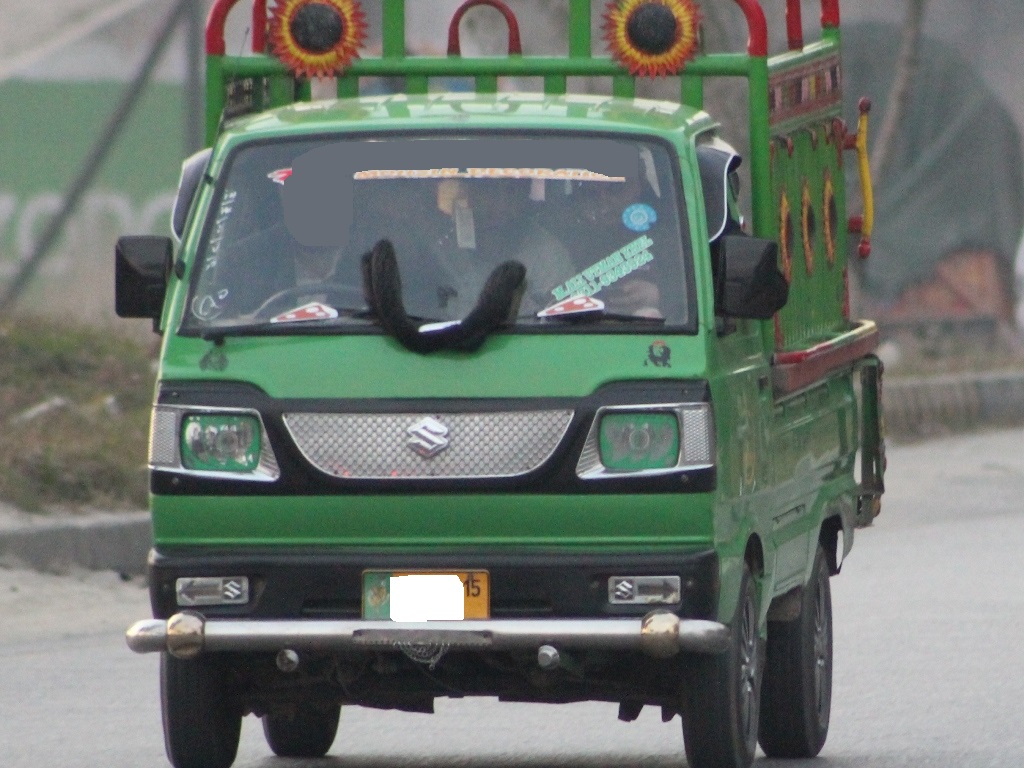}&
\includegraphics[width=.15\textwidth]{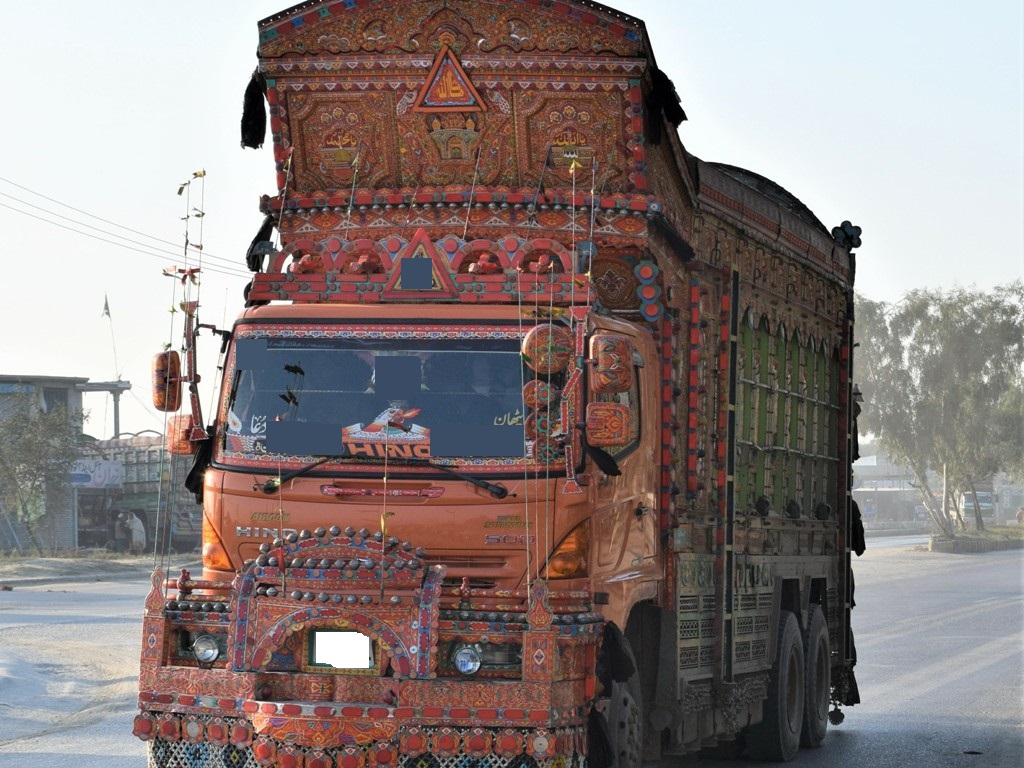}&
\includegraphics[width=.15\textwidth]{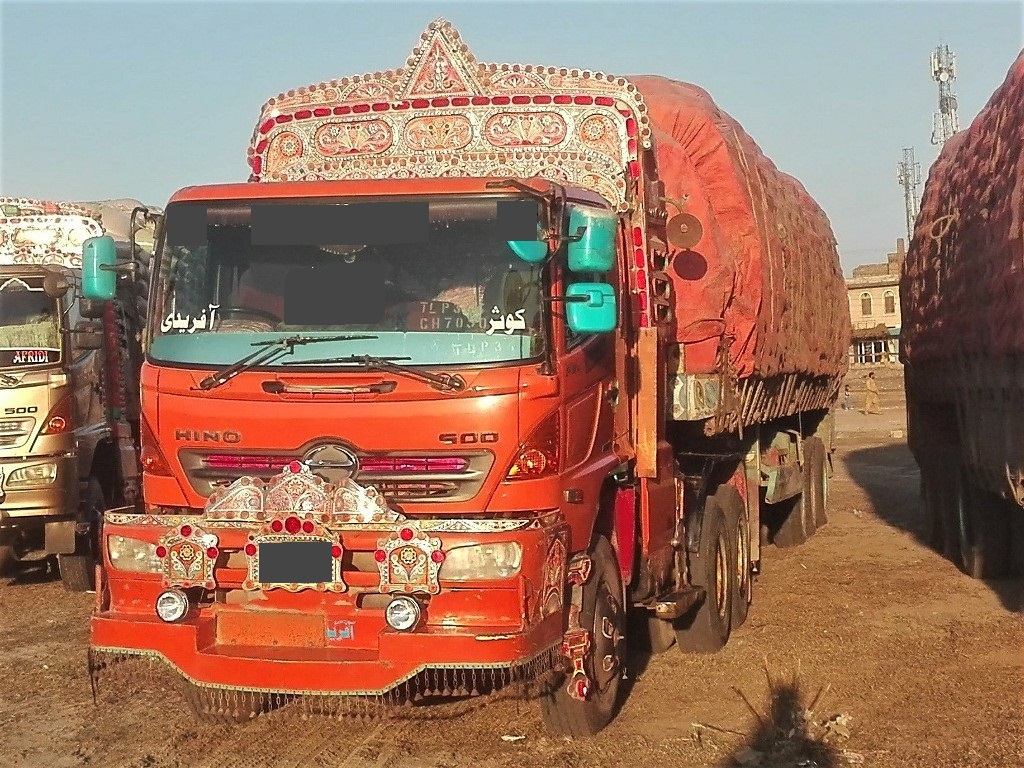}&
\includegraphics[width=.15\textwidth]{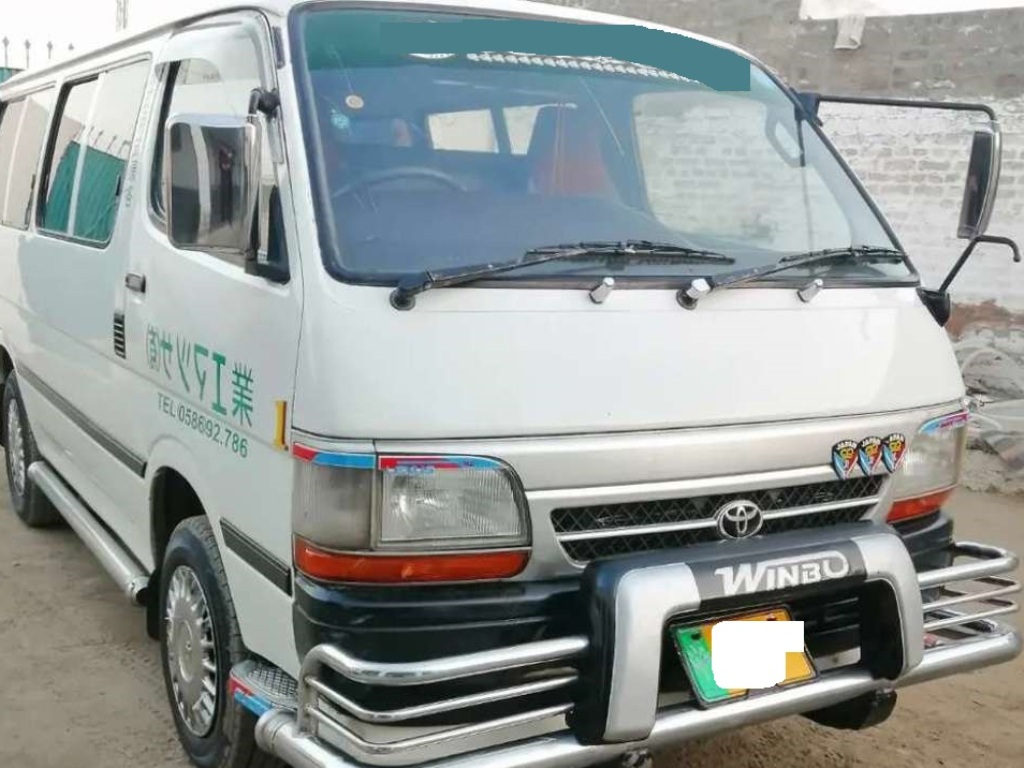}\\

\end{tabular}
\end{center}
\caption{Exemplar images of vehicle types. The intra-class variations can be observed due to decorations, manufacturers and other shape modifications such bumpers and steel bodies for goods transport}
\label{fig:3}
\end{figure*}

\textbf{CNN-based methods}:
The CNN-based framework~\cite{jamtsho2020real} is proposed in which, before localizing LP, a vehicle is detected to eliminate false positives and other objects such as signboards, \etc, employing CNN-based famous object detector framework YOLO~\cite{redmon2018yolov3} to localize the LP. 
Various versions of Bangladeshi license plates~\cite{onim2020traffic} are detected using a modified version of YOLO called YOLOv4, employing Tesseract~\cite{TessOverview}  as an OCR engine to recognize characters from the detected license plates. 
Mask RCNN~\cite{8237584} is used for the detection, segmentation, and recognition of Tunisian license plates~\cite{selmi2020delp} under challenging image variations caused by environmental conditions, cluttered background, orientation, and language differences.

Yet another CNN-based method~\cite{bjorklund2019robust} is proposed for the detection and recognition of license plates. However, unlike most techniques, the authors relied on synthetically generated license plate images that suffered from the common variations found in the real license plate images. Consequently, showing that their framework outperformed the CNN trained on natural images of the license plates.  
Yang~\etal~\cite{yang2019recognizing} propose a real-time coarse to fine strategy via contours detection and reconstruction for license plate detection while using a CNN for final character recognition. 
A layout-independent method\cite{laroca2019efficient} is proposed for detecting and recognizing the license plates that belong to several countries. To this end, as a first step, detecting vehicle(s) in a given image and then detect and read the license plates. Both the steps are performed using the YOLO V3 object localization framework.

Wang~\etal\cite{wang2020rethinking} propose a two-step process for license plate detection and recognition by designing two specialized networks. For detection, VertexNet is designed to extract the spatial information of the license plate and later on uses this information to rectify the detected license plate image. For character recognition, the detected license plate is then fed to the second network named SCR-Net. The semantic segmentation based on DeepLabv2~\cite{chen2017deeplab} is utilized by~~\cite{zhuang2018towards} to extract the license plate from the image, which is then followed by counting the number of segments that are candidates of license plate characters. Such segmentation and counting make their method fast and accurate as of the individual steps of license plate detection, extraction, and recognition are bypassed. Other CNN-based methods for license plate detection and recognition use YOLO~\cite{tourani2020robust}, LSTM~\cite{li2016reading}, and Attentional Networks~\cite{zhang2020robust}. 
Nonetheless, we aim at license plates detection and recognition such that, unlike most previous methods, their locations on the vehicles, layout, and font styles are not standardized. 

\section{Dataset}
\label{sec:dataset}
\begin{figure*}[t!]
    \centering
    \includegraphics[width=1\textwidth]{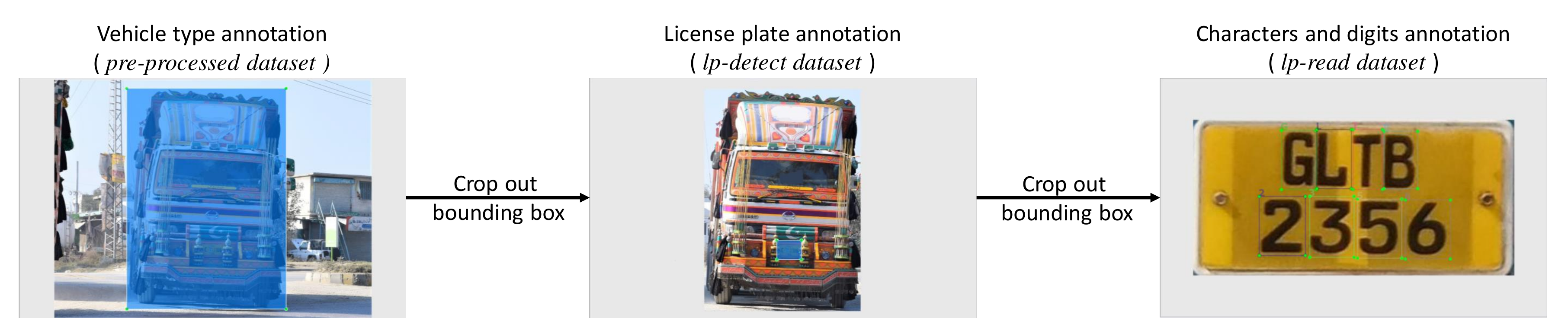}
    \caption{Ground truth generation process the three steps.}
    \label{fig:4}
\end{figure*}

We collect a novel dataset,\emph{Diverse Vehicle and License Plates Dataset (DVLPD)}, that has 10k images of Pakistani vehicles. These vehicles are trucks, buses, vans, carry vans, and cars. There are two sub-classes of trucks called \emph{type1} trucks having a single axle and \emph{type2} trucks that have double axles. Figure~\ref{fig:3} shows the exemplar images for each vehicle. Since our approach to the problem is data-driven, we carefully accommodate all the challenges mentioned above in the images of each vehicle type. After dataset collection, the pre-processing steps and ground truth generation are performed to train and test the CNN models. These steps are explained subsequently and are shown in Figure~\ref{fig:4}, whereas the statistics about vehicles images and their license plates are given in Figure~\ref{fig:5}. 
\subsection{Data pre-processing}
 
\begin{enumerate}
    \item \textbf{Cropping}: The images are cropped with a standard aspect ratio of 3:4 to remove the unwanted information. For instance, a vehicle image taken from a long distance contains a patch of road that acts as redundant information. Apart from this, cropping also reduces the image size, which is important for the reduced training time of the deep learning model. 
    \item \textbf{Resizing}: The images are resized because the network must receive the same size images during the training process. The image size depends greatly on the problem being dealt with and the architecture’s requirement. Hence, there is no standard defined for the size of the input images; it can be of different resolutions, such as $ 416 \times 416$ and $ 256 \times 256$. Images with lower resolution require less processing time, but the trade-off is fewer details in the image. In this research, the image is resized to $ 416 \times 416$ pixels for the training and testing. Most architectures recommend the chosen size because the details are preserved without losing important information in the images.
\end{enumerate}

\subsection{Ground truth generation}
\begin{figure}
     \centering
     \begin{subfigure}[b]{1\columnwidth}
         \centering
         \subcaption[short for lof]{Number of instances for vehicle types in \textit{pre-processed} and \textit{lp-detect dataset}}
         \includegraphics[width=\textwidth]{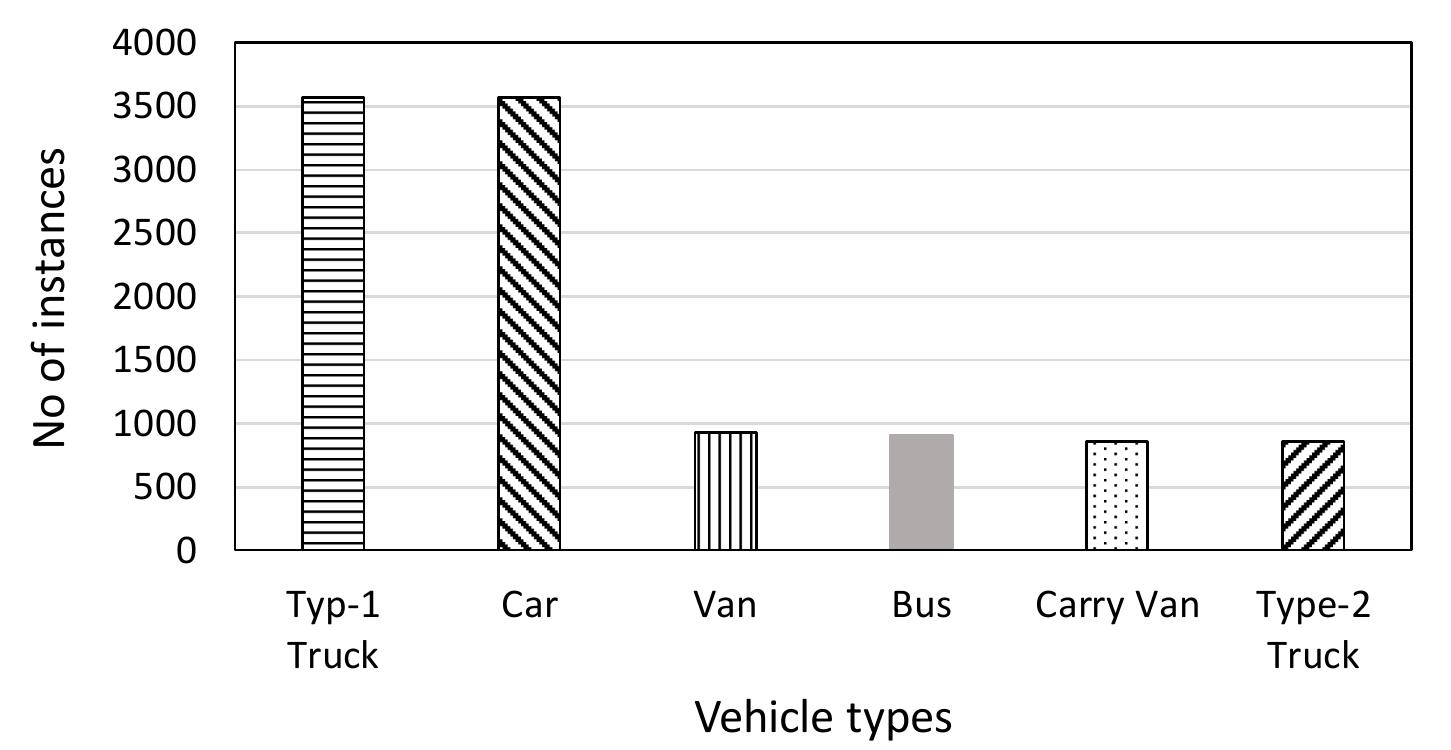}
         \label{fig:411}
     \end{subfigure}
     \hfill
     \begin{subfigure}[b]{1\columnwidth}
         \centering
         \subcaption[short for lof]{Number of instances for each digit and character in \textit{lp-read dataset}}
         \includegraphics[width=\textwidth]{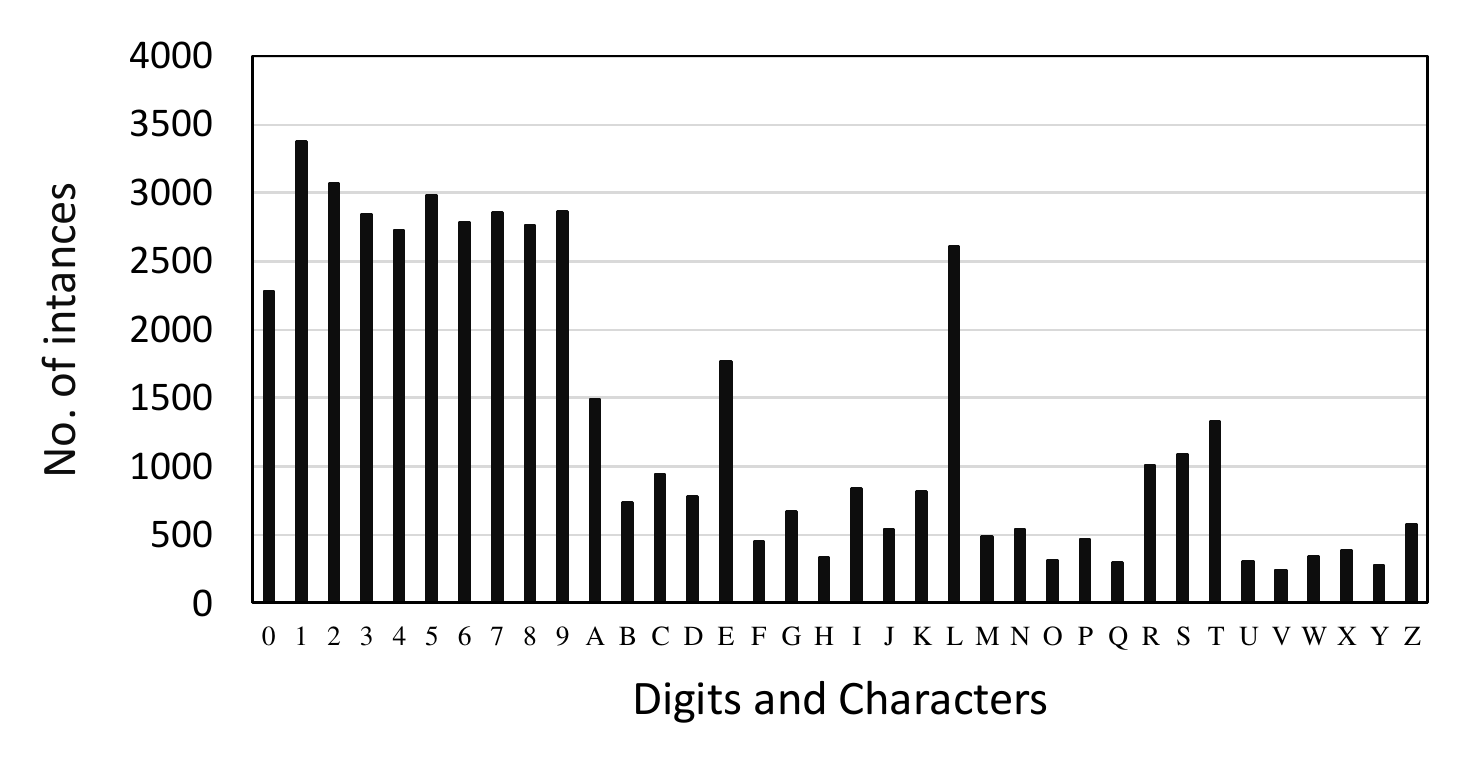}
         \label{fig:412}
     \end{subfigure}
    \caption{Statistics of the \textit{pre-processed}, \textit{lp-detect dataset}, and \textit{lp-read dataset dataset}}
        \label{fig:5}
\end{figure}

As mentioned previously, we train a separate model for each of the three steps \ie, vehicle type recognition, license plate detection, and license plate character recognition. Consequently, all the vehicle images are annotated for each of the individual steps in the following manner. 
\begin{enumerate}
    \item \textbf{Vehicle type detection}: First, we annotate all the pre-processed images for six vehicle types using the LabelIMG tool~\cite{tzutalin2015labelimg}. The ground truth bounding box is manually drawn only around the depicted vehicle in the image. This labeled dataset is employed to train and test the model for vehicle type recognition.  
    \item \textbf{License plate detection}: Subsequently, the ground truth bounding boxes around the vehicles are used to crop out all the depicted vehicles. This generates a sub-dataset which we call \enquote{\textit{lp-detect dataset}}. The dataset images are then prepared for the second step of our framework, which is license plate detection. The bounding boxes are manually drawn around the license plates of the vehicles. These annotated images are utilized for training and testing the second model for license plate detection.
    \item \textbf{License plate reading}: For the final step of license plates reading, we make another dataset by cropping out all the license plates from the \textit{lp-detect dataset dataset} which is named as \textit{lp-read dataset dataset}. This dataset is then annotated for digits (0-9) and characters (A-Z) for the final step of license plates reading.
\end{enumerate}
The statistics of each of the datasets are shown in Figure~\ref{fig:5}.

\section{Methodology}
\label{sec:method}
The proposed framework for the image-based toll tax calculation is explained in this section. The toll tax is calculated by designing a framework that performs image-based end-to-end recognition of the vehicle’s type and license plate. Due to the challenging image variations, we have focused on developing a system that uses three object detection models. With such a strategy, we aim to achieve the highest possible accuracy at recognizing the type of the vehicle, localize its license plate, and finally identify the license plate’s characters. Each of these steps is explained in the following and their deployment on a Raspberry Pi for real-time usage. 

\subsection{Proposed Model}
\begin{figure}[!t]
    \centering
    \includegraphics[width=1\columnwidth]{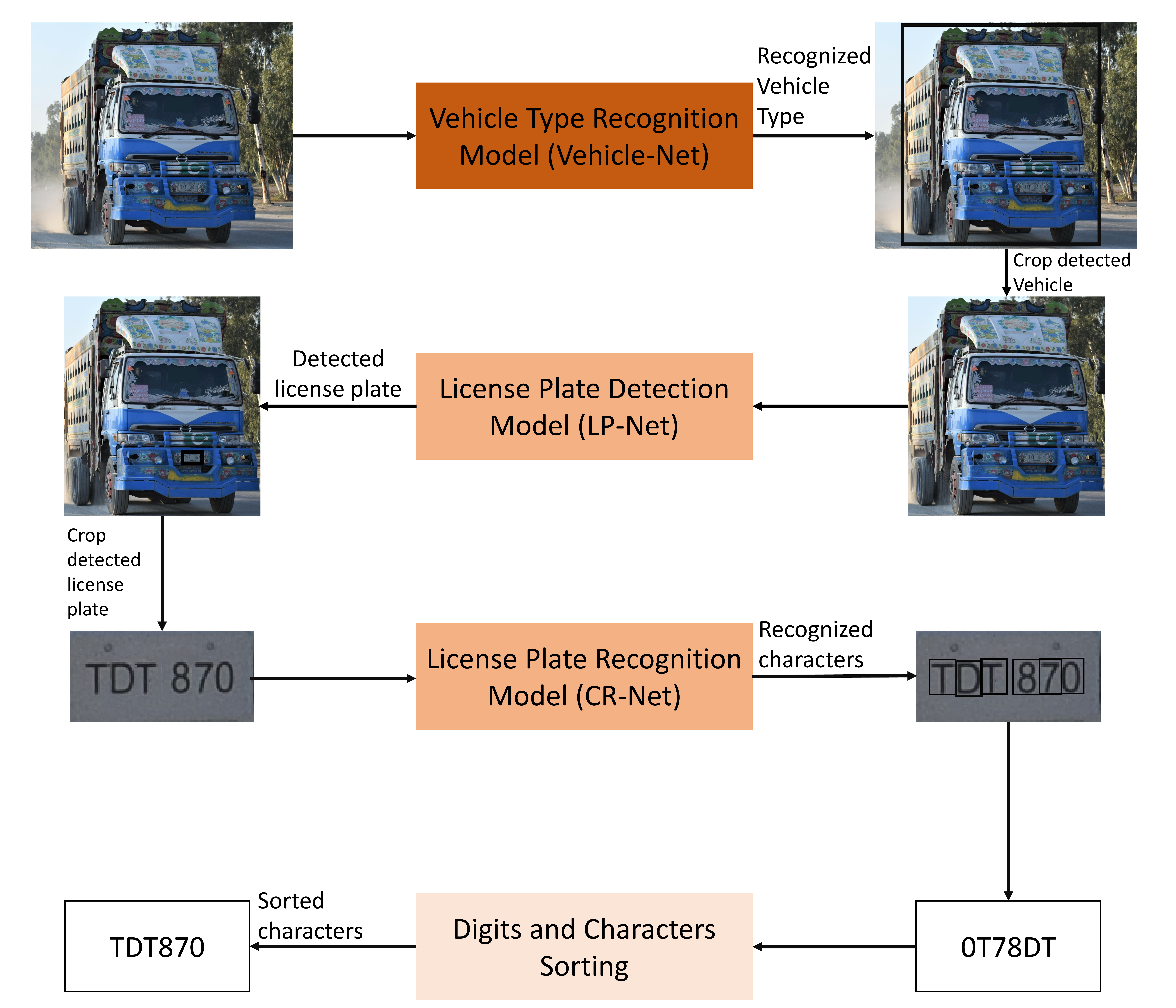}
    \caption{Block Diagram showing the full working for our proposed framework. The different colors represent a specific module in our framework.}
    \label{fig:6}
\end{figure}

Figure~\ref{fig:6} shows the block diagram of the entire framework. The input to this framework can be an image, series of images, or video frames, whereas the outputs are the recognized vehicle type and license plate characters. When an image is fed to the framework, the first object detection model is triggered for the vehicle type recognition, which we call the \enquote{\textit{Vehicle-Net}.} The vehicle is localized in the image via a bounding box, and its type is predicted. 
The coordinates of this bounding box are then used to extract the vehicle, and the next model for license plate detection called the \enquote{\textit{LP-Net}} is applied to it. This model predicts the bounding box around the license plate of the vehicle. Finally, the last model called the \enquote{CR-Net} is then triggered on the detected license plate, which specializes in detecting and recognizing the characters of the license plate. In such a way, each input frame is processed seamlessly by the three models, with a final output giving the labels of the detected type of vehicle and the characters in its license plate. When there is no vehicle present in the input frame at the first object detection network, the frame is skipped, and the system waits for the next frame to detect a vehicle in the image.

\subsection{RaspberryPi Deployment}
\begin{figure}[t!]
    \centering
    \includegraphics[width=1\columnwidth]{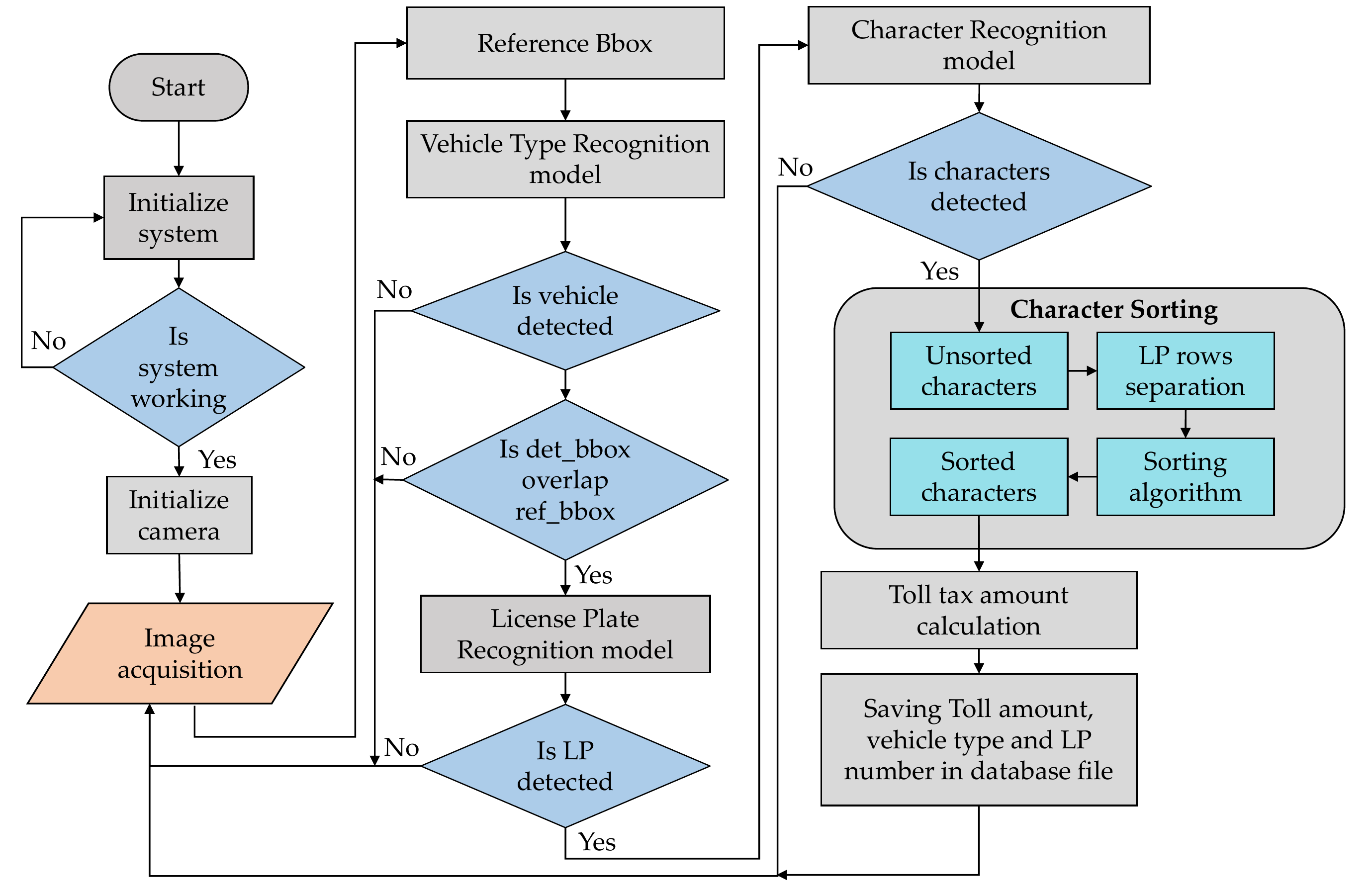}
    \caption{Flow Process of Proposed Methodology}
    \label{fig:7}
\end{figure}

After the offline training, all three models are deployed on a RaspberryPi 4 (RPi). As shown in Figure~\ref{fig:5}, after acquiring images with the help of a Pi camera interfaced at the camera port, the device applies all three models on them for a vehicle type and license plate recognition. 
Figure~\ref{fig:7} shows the flowchart that depicts the technical working of the algorithm running on the RPi. As a first step, the models are loaded into the memory of RPi. The RPi has a RAM of 2GB which is sufficient to load the tiny models of YOLOv3 and YOLOv4. In the next step, the camera is initialized, and a video stream is an input into the models’ frame by frame. In the next step, a reference bounding box is drawn to tackle the problem of multiple detections, which is explained in Section~\ref{multipleVehicles}. After the vehicle is detected, the output of the first network is cropped out using the coordinates of the bounding box with the highest confidence score. This output is then fed into the next model to detect the license plate in the frame. Similarly, the frame is processed again from the network's final layer, and this time the license plate is fed into the last network, which recognizes the characters in the image. The recognized characters are then sorted according to their detected bounding boxes using the sorting algorithm, which is explained in Section~\ref{characterSorting}.
\begin{algorithm}[t]
\caption{Multiple Vehicle Detection}\label{alg:cap}
\begin{algorithmic}
\label{alg:Alg1}
\Require \text{Draw Reference Bounding Box on camera frame}
\Ensure $Input Frame \geq 0$
\State $RefBBox[\, \,] \gets \textit{xmin, ymin, xmax, ymax}$
\State $indices \gets \textit{Detection Results}$
\For {\texttt{i in indices}}
\If{$len(indices)$ is > 0}
    \State $left\gets \textit{Detected BBox Left Coordinate}$
    \State $right \gets \textit{Detected BBox Right Coordinate}$
    \State $top \gets \textit{Detected BBox Top Coordinate}$
    \State $bottom \gets \textit{Detected BBox Bottom Coordinate}$

\State $DetBBox[\, \,] \gets \textit{xmin, ymin, xmax, ymax}$

\State $xA \gets (max(RefBBox[0],DetBBox[0]))$
\newline
\Comment{Computing max and min of RefBBox[\,\,] and DetBBox[\,\,]}
\State $yA \gets (max(RefBBox[1],DetBBox[1]))$
\State $xB \gets (max(RefBBox[2],DetBBox[2]))$
\State $yB \gets (max(RefBBox[3],DetBBox[3]))$
\State $overlapArea \gets abs(max((xB-xA,0)) * max((yB-yA),0))$

\If{$overlapArea \leq 0 $}
     \State Overlap Exists
 \Else
 \State No Overlap Exists
\EndIf
\EndIf
\EndFor
\end{algorithmic}
\end{algorithm}
\subsubsection{Multiple Vehicle Detection}
\label{multipleVehicles}
Multiple vehicle detection in a single image makes it difficult to select the vehicle for which toll tax calculation must be performed. The camera position on the toll tax collection plaza gives a rough estimate of the expected position of the vehicle in the image region. We draw a so-called \enquote{\textit{reference box}} around this image region and store its coordinates. Once a vehicle is detected, the algorithm iterates over all the bounding boxes detected by the network and checks whether the overlap area between the reference and detected bounding box exists using a formula on the coordinates of both reference and detection. The detailed algorithm for avoiding multiple vehicle detections is outlined in Algorithm~\ref{alg:Alg1}, where the coordinates of the reference bounding box are compared with the detection results.
\subsubsection{Character Sorting in License Plate}
\label{characterSorting}
\begin{figure}[t]
    \centering
    \includegraphics[width=1\columnwidth]{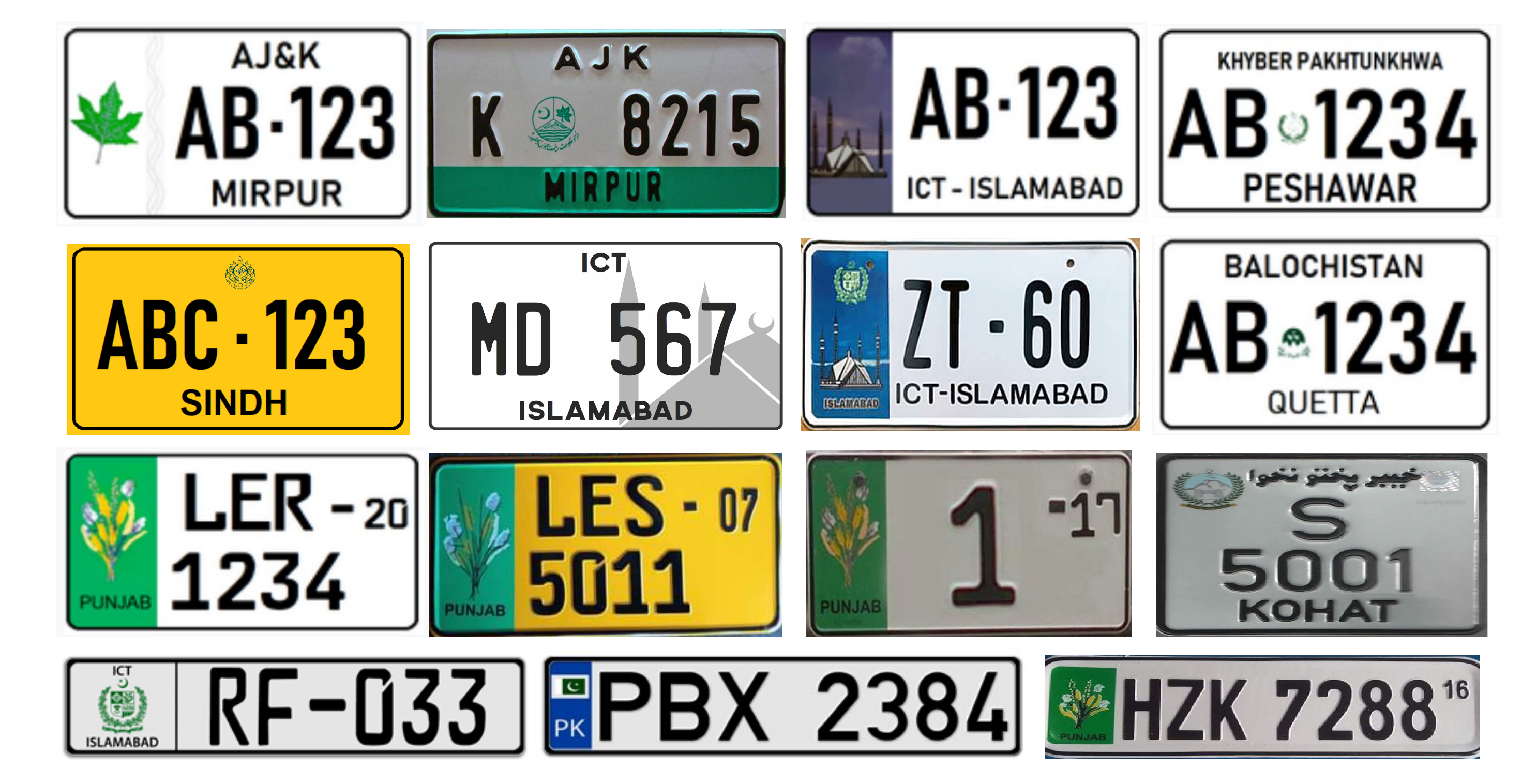}
    \caption{Different types of license plates showing the diversity in style and arrangement.}
    \label{fig:s}
\end{figure}

The sequence of the recognized characters and digits is not identical to that of the license plate due to the fact that \textit{CR-Net} does not sort its output. While dealing with license plates, the sequence of the characters is vital. The characters of license plates are usually arranged in one or two rows, as shown in Figure~\ref{fig:s}. To sort the characters in more than one row, we use the detected labels and the coordinates of their bounding boxes to sequence the characters in the license plate correctly. The labels and the coordinates of those labels are stored, then the bounding boxes with labels as alphabets and numbers are separated. Then, the coordinates of the bounding boxes in the alphabets and numbers are sorted using the \enquote{x-min} coordinate from left to right in detected bounding boxes and displayed as the final correctly sequenced license plate.
\subsection{Backbone Architectures}
Various State-of-the-art (SOTA) architectures are used as the backbone model in our proposed methodology to accomplish vehicle type recognition, license plate extraction, and character recognition tasks. 
\begin{enumerate}
    \item \textbf{Faster RCNN}:
    Faster RCNN uses Deep CNN called Region proposal Network (RPN) to generate the sets of region proposal bounding boxes. After the region proposals are obtained, they are subjected to another network for extracting the proposal features from the feature map and classification using softmax. This belongs to the RCNN family, which is a multi-stage object detection algorithm due to which the real-time performance is barely reachable.
    \item \textbf{YOLOv2}:
    Single staged end-to-end training and detection are achieved through YOLO. This one-stage detector treats object detection as a regression problem, thus closing the gap of real-time performance. YOLOv2 framework is called Darknet-19 which consists of 19 convolutional layers and five max-pooling layers. It uses multi-resolution images by removing fully connected layers. Global average pooling is used for the prediction, and batch normalization is introduced after each convolution layer to regularize the model and speed up the convergence. 
    \item \textbf{YOLOv3 \& Tiny YOLOv3}:
    YOLOv3 is more optimized in terms of detection speed and accuracy as compared to previous models. Feature Pyramid Network (FPN) approach is introduced in YOLOv3, which allows detecting the objects at three different scales. It also uses residual blocks to tackle the problem of vanishing gradients in a dense network.
    The YOLOv3 tiny is a relatively lightweight architecture as compared to the YOLOv3. It is utilized for an embedded system where the resources are limited. The detection accuracy of the lighter architecture is less than the complete architecture, while the performance in FPS for Tiny YOLOv3 is greater than YOLOv3. The feature extractor is the main difference in both architectures, where YOLOv3 uses the Darknet-53 with 53 layers, and the Tiny YOLOv3 uses a compressed version with seven convolutional layers and six max-pooling layers in the feature extractor.
    
    \item \textbf{YOLOv4 and Tiny YOLOv4}:
    YOLO v4 is introduced with numerous improvements, in particular with Path Aggregation Network (PAN). It is used to communicate information from lower layers to higher ones. In YOLO v4, the CSPDARKNET53 model is used as the backbone, a CNN that uses DarkNet-53. It employs a split and merges strategy to split the feature map of the base layer into two parts and then merge them. This allows for more gradient flow through the network. Spatial Pyramid Pooling (SPP) and PAN are employed as the neck and YOLO v3 network as the head. A new data augmentation technique is applied called Mosaic data augmentation. In this technique, four images of different aspect ratios are combined into a single image, which helps the model to find smaller objects and pay less focus to the surrounding scenes.
    Tiny YOLOv4 is a lighter and compressed version of YOLOv4. Tiny YOLOv4 has a simpler architecture that contains 29 convolutional layers, whereas YOLOv4 contains 137 pre-trained convolutional layers; thus, the tiny version is more feasible for embedded and mobile applications. Tiny YOLOv4 is eight times faster in terms of Frame Per Second (FPS) than YOLOv4, whereas accuracy for tiny version is 2/3rds of YOLOv4.
\end{enumerate}

\section{Result and discussion}
\label{sec:results}
The experimental results are discussed in this section systematically. In the first place, we outline the protocols and hardware used in the experiments. We then give some attributes of the datasets used in the training and testing of each model. The results for vehicle type recognition, license plate detection, and license plate recognition are then reported in the form of precision, recall, F1-score, and mean average precision (mAP). We then report an end-to-end accuracy achieved by our framework, and finally, the execution time of the backbone architectures on Raspberry Pi is reported. 
\subsection{Experimental Protocols}
For the training and testing of each model, we randomly split their related image dataset into 80\% training and 20\% test set and recorded their results. Since the split is random, we carry out such training and testing five times and then report their average results and standard deviation. The hardware system specifications for training and testing are shown in Table~\ref{tab:my-table1}.

We use the Darknet framework to implement YOLOv4, Tiny YOLOv4, YOLOv3, Tiny YOLOv3,  YOLOv2, and the TensorFlow Object Detection API for faster RCNN implementation. The trained models are tested with the color images of resolution $416 \times 416$ pixels. The YOLO model has three scale pyramids used to detect small, medium, and large objects. The number of epochs for training Vehicle-Net, LP-Net, and CR-Net is 8k, 2k, and 10K, respectively. The batch size is 64, where the subdivisions are 32. The activation function used in YOLOv3, Tiny YOLOv3, and YOLO v2 is Leaky ReLU, YOLOv4, and Tiny YOLOv4 use the Mish activation while FasterRCNN uses the Relu activation function. The training loss depends on the losses caused by objectness, classification, and coordinate loss. The learning rate is fixed at $10^{-3}$. Augmentation is also applied to the training images to accommodate for the variations caused due to saturation, hue, and exposure. The threshold for non-maxima suppression (NMS) is set to 0.3 while the IOU threshold is set to 0.5, which means that the detection is considered correct if the overlap is greater than the confidence score of 0.5. All the parameters are summarized in Table~\ref{tab:my-table2}. 
\begin{table}[t!]
\caption{Hardware Specifications for training the detection models}
\label{tab:my-table1}
\begin{center}
\begin{tabular}{ll}
\toprule
\textbf{Resource Type} & \textbf{Specification}  \\ \midrule
CPU                    & Intel Core i7-6700      \\
RAM                    & 16GB DDR 4              \\
GPU                    & Nvidia GeForce GTX 1060 \\
CPU Cores              & 4                       \\
Frequency              & 3.4 GHz                 \\ \bottomrule
\end{tabular}
\end{center}
\end{table}
\begin{table}[t!]
\caption{Training parameters of detection models}
\label{tab:my-table2}
\begin{center}
\begin{tabular}{l l}
\toprule
\textbf{Parameters}\ & \textbf{Configuration} \\ \midrule
Framework                            & Darknet \& Tensorflow object detection API               \\
Max Batches                          & 3k, 5k \& 10k                          \\
Batch                                & 64                                      \\
Subdivisions                         & 32                                      \\
Width                                & 416                                     \\
Height                               & 416                                     \\
Channels                             & 3 (RGB)                                 \\
Policy                               & Steps                                   \\
Learning rate                        & 0.001                                   \\
Momentum                             & 0.9                                     \\
Decay                                & 0.0005                                  \\
Saturation                           & 1.5                                     \\
Exposure                             & 1.5                                     \\
Hue                                  & 0.1                                     \\
NMS Threshold                        & 0.3                                     \\ \bottomrule
\end{tabular}
\end{center}
\end{table}
\subsection{Dataset Attributes}
Our three image datasets are \textit{pre-processed dataset} for Vehicle-Net, \textit{lp-detect dataset} for LP-Net and \textit{lp-read} dataset for the CR-Net. The \textit{pre-processed dataset} training and testing sets contain 7,788 and 1,947 images, respectively, while the total vehicle annotation in this dataset is 10,218. In the \textit{lp-detect dataset} the training and test sets contain 7,891 and 1,972 images, respectively, while the dataset has 9,935 annotated license plates of each vehicle. Lastly, the \textit{lp-read} dataset, which is used to recognize license plate characters, contains 6,261 images in the training set and 1,565 images in the test set. The total number of annotated characters in the dataset is 47,711. 
\begin{figure}[t!]
     \centering
    \includegraphics[width=1\columnwidth]{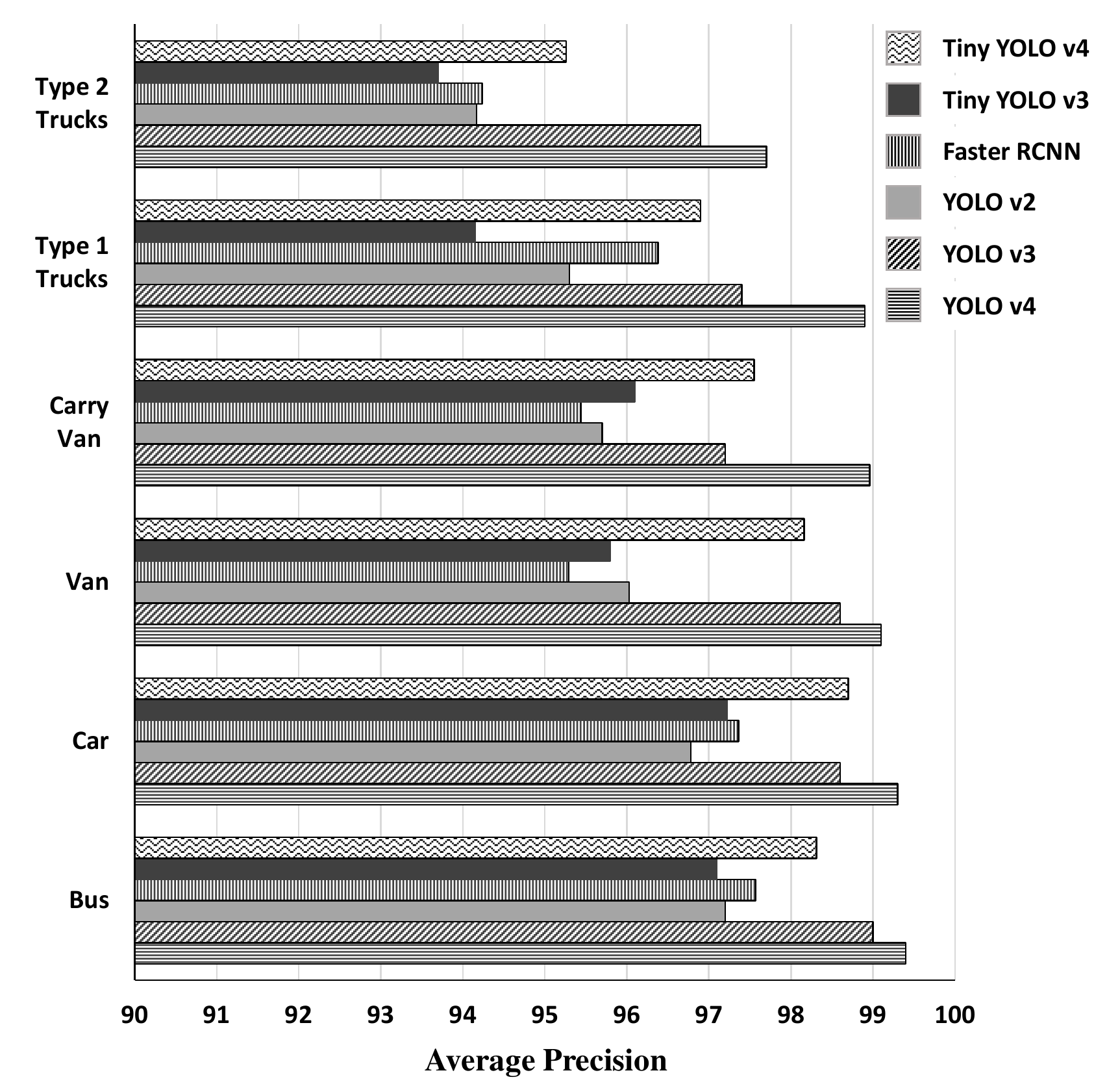}
     \caption{Average precision achieved for each of the vehicle types}
     \label{fig:8}
\end{figure}
\begin{table}[t]
\caption{Results  of  vehicle  type  recognition  achieved  by various architectures}
\label{tab:type-recognition}
\begin{tabularx}{\columnwidth}{L{1.9cm} C{1.2cm} C{1.2cm} C{1.2cm} C{1.2cm}}
\toprule
\multicolumn{5}{c}{Vehicle Type Recognition}                                                                                                \\ \midrule
                      & \multicolumn{1}{c}{\textbf{Precision}} & \multicolumn{1}{c}{\textbf{Recall}} & \multicolumn{1}{c}{\textbf{F1-score}} & \multicolumn{1}{c}{\textbf{mAP@0.5}} \\ \midrule

\textbf{Tiny YOLOv4}  & 95.8$\pm$1.2  & 96.4$\pm$0.5 & 95.8$\pm$0.7                        & 97.1$\pm$0.4       \\

\textbf{Tiny YOLOv3}  & 92.0$\pm$2.3  & 92.2$\pm$1.7 & 92.0$\pm$1.4                        & 94.8$\pm$1.0       \\

\textbf{Faster-RCNN}  & 96.0$\pm$0.6  & 95.8$\pm$1.9 & 95.6$\pm$1.0                       & 96.1$\pm$0.6       \\

\textbf{YOLOv2}       & 92.4$\pm$1.3  & 92.2$\pm$2.3 & 93.2$\pm$0.7                        & 95.0$\pm$0.6       \\

\textbf{YOLOv3}       & 95.6$\pm$1.8  & 97.6$\pm$0.5 & 96.4$\pm$1.0                        & 97.6$\pm$0.7       \\

\textbf{YOLOv4}       & 98.4$\pm$0.8  & 97.8$\pm$0.4 & 98.2$\pm$0.7                        & 98.8$\pm$0.4      \\ \bottomrule
\end{tabularx}
\end{table}
\begin{figure}[t!]
    \centering
    \includegraphics[width=1\columnwidth]{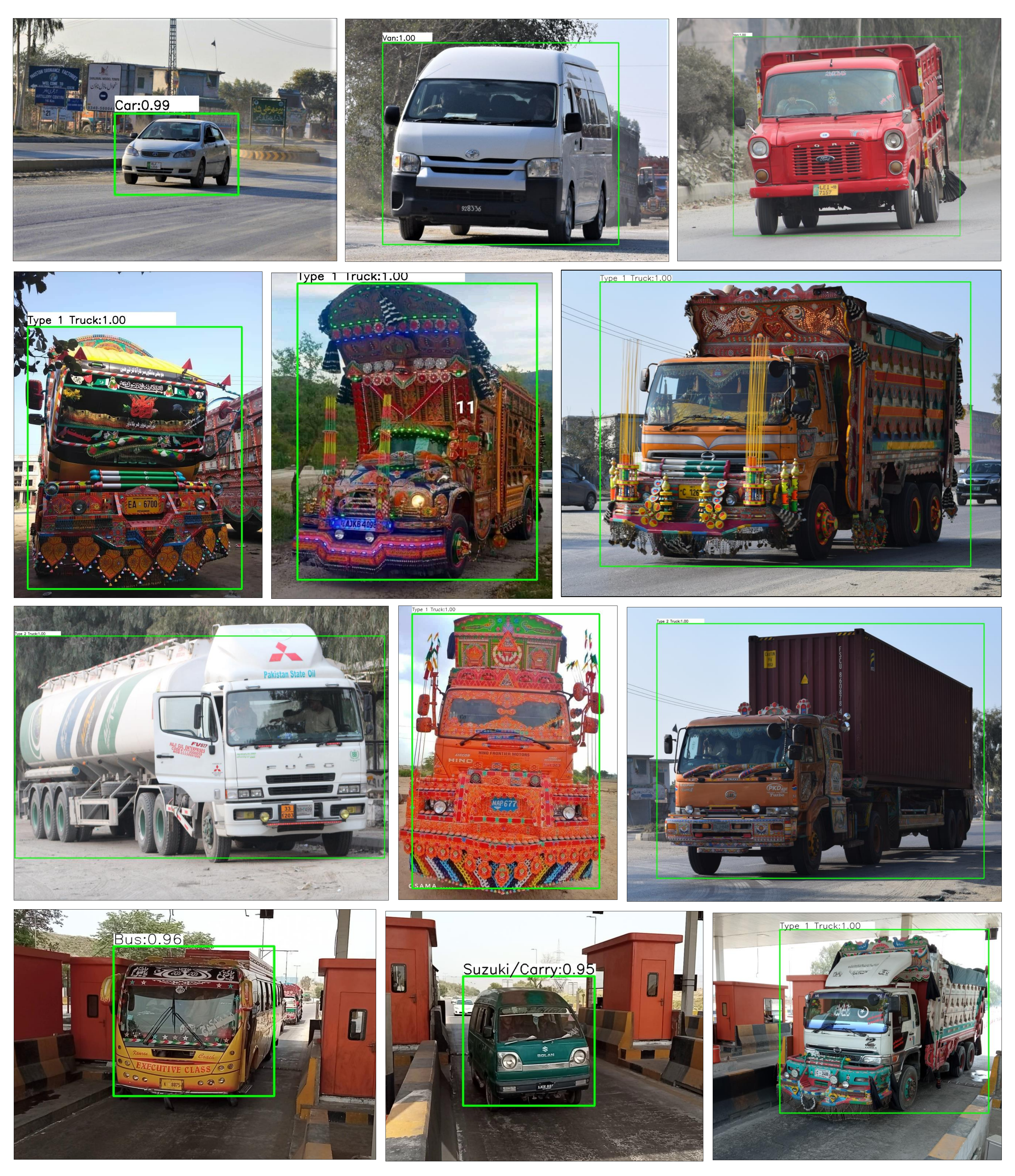}
    \caption{Vehicle type recognition. The last row shows the detection results of the images taken at the toll plaza}
    \label{fig:9}
\end{figure}
\subsection{Vehicle Type Recognition}
The average precision achieved by each architecture for each of the vehicle types is shown in Figure~\ref{fig:8}. YOLOv4 achieves the highest scores for all the types, while its tiny version performs better than the tiny version of YOLOv3. As mentioned before, due to random shuffling of the dataset, the experiments are performed five times where for each experiment, the number of training iterations is 8K. 
It can be noted that the best result of 99\% is achieved for buses, cars, and vans, while trucks are recognized the least. We believe that this is mainly due to the high intra-class variations in trucks due to the heavy decorations. The least precision is achieved for type-2 trucks, mainly due to their similarity with type-1 trucks due to front decorations and models from similar manufacturers. Nonetheless, YOLOv4 achieves the best mAP of 98.4\% for vehicle type recognition followed by YOLOv3 and Tiny YOLOv4 as shown in Table~\ref{tab:type-recognition}. Some vehicle detection results are shown in Figure~\ref{fig:9} where the last row shows the detected vehicle on a toll collection plaza. 
\begin{figure}[b!]
     \centering
    \includegraphics[width=1\columnwidth]{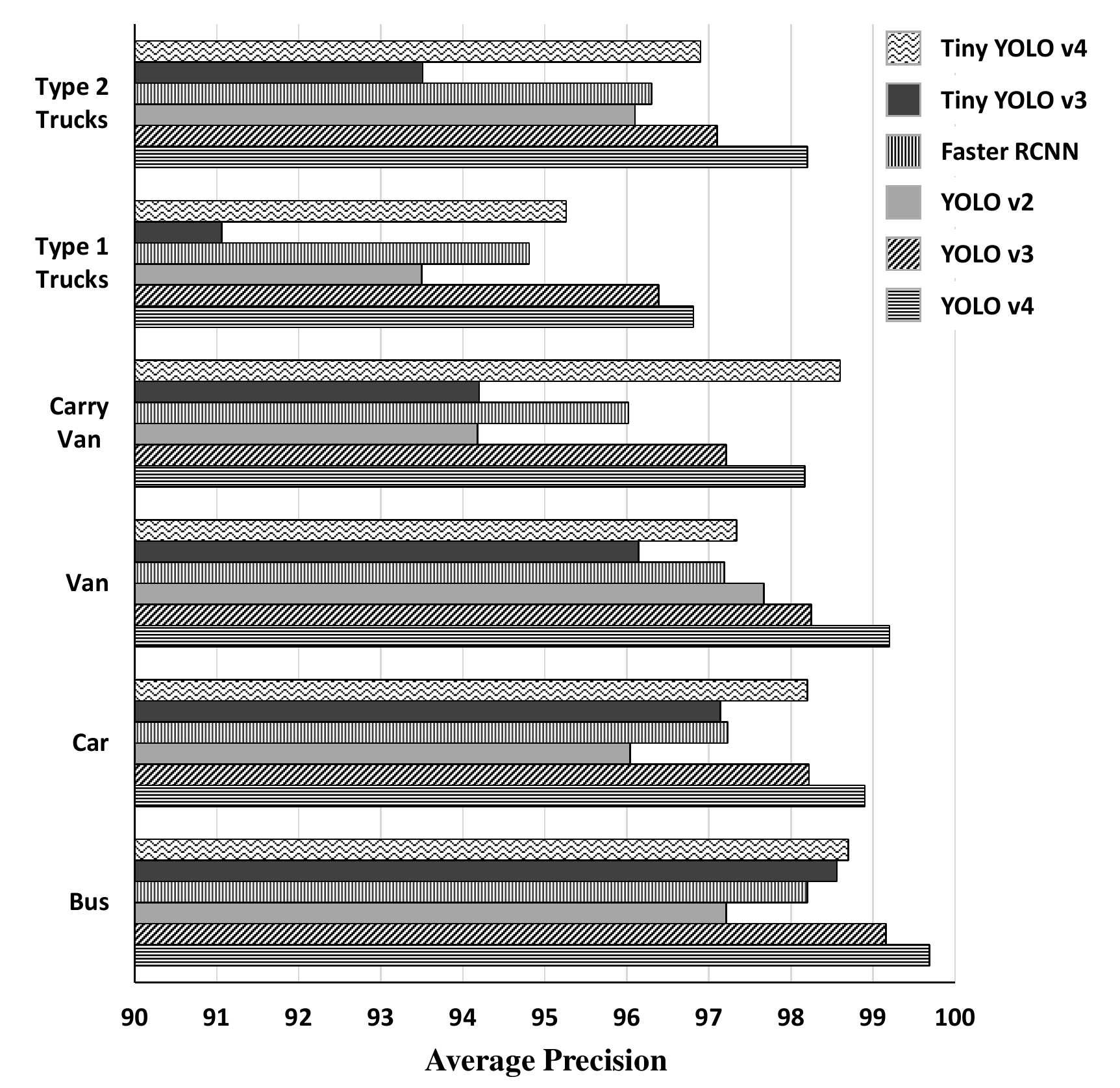}
     \caption{Average precision achieved by various architectures for the detection of license plates at each of the vehicle types}
     \label{fig:10}
\end{figure}
\begin{table}[t]
\caption{Quantitative performance of state-of-the-art methods for license  plate  detection.}
\label{tab:plate-detect}
\begin{tabularx}{\columnwidth}{L{1.9cm} C{1.2cm} C{1.2cm} C{1.2cm} C{1.2cm}}
\toprule
\multicolumn{5}{c}{\textbf{License Plate Detection}}                                                                                                                            \\ \midrule
                      & \multicolumn{1}{c}{\textbf{Precision}} & \multicolumn{1}{c}{\textbf{Recall}} & \multicolumn{1}{c}{\textbf{F1-score}} & \multicolumn{1}{c}{\textbf{mAP@0.5}} \\ \midrule

\textbf{Tiny YOLOv4}    & 96.6$\pm$1.0 & 97.6$\pm$1.0 & 96.8$\pm$0.7                             & 97.0$\pm$0.5                          \\

\textbf{Tiny YOLOv3}    & 93.0$\pm$1.4 & 92.0$\pm$1.4 & 92.6$\pm$1.0                             & 95.0$\pm$0.7                       \\

\textbf{Faster-RCNN}    & 96.2$\pm$1.3 & 96.4$\pm$1.0 & 96.2$\pm$0.7                             & 96.0$\pm$0.7                          \\

\textbf{YOLOv2}         & 93.4$\pm$1.3 & 93.0$\pm$1.7 & 93.4$\pm$1.0                             & 95.3$\pm$0.8                        \\

\textbf{YOLOv3}         & 97.6$\pm$0.5 & 97.6$\pm$0.5 & 97.6$\pm$0.5                             & 98.0$\pm$0.4                          \\

\textbf{YOLOv4}         & 97.4$\pm$1.0 & 97.4$\pm$1.2 & 97.4$\pm$0.5                             & 98.5$\pm$0.3                         \\ \bottomrule
\end{tabularx}
\end{table}

\subsection{License plate detection}
License plate detection is the most crucial step in our proposed framework. This is since the Pakistani vehicles are heavily decorated and exposed to all sorts of weather conditions. Over time, license plates are covered with dust and mud, which makes their detection challenging. Like the vehicle type recognition, we perform the experiments five times, whereby for each experiment, the \textit{lp-detect} is randomly split into training and test sets. Each time, the number of training iterations are 2K. Figure~\ref{fig:10} shows the average precision achieved by each architecture on the license plates of all the vehicle types, while the detailed results are shown in Table~\ref{tab:plate-detect}. The license plates of buses are most accurately detected, while those on the type-1 trucks are detected the least. 
\begin{figure}[t]
    \centering
    \includegraphics[width=1\columnwidth]{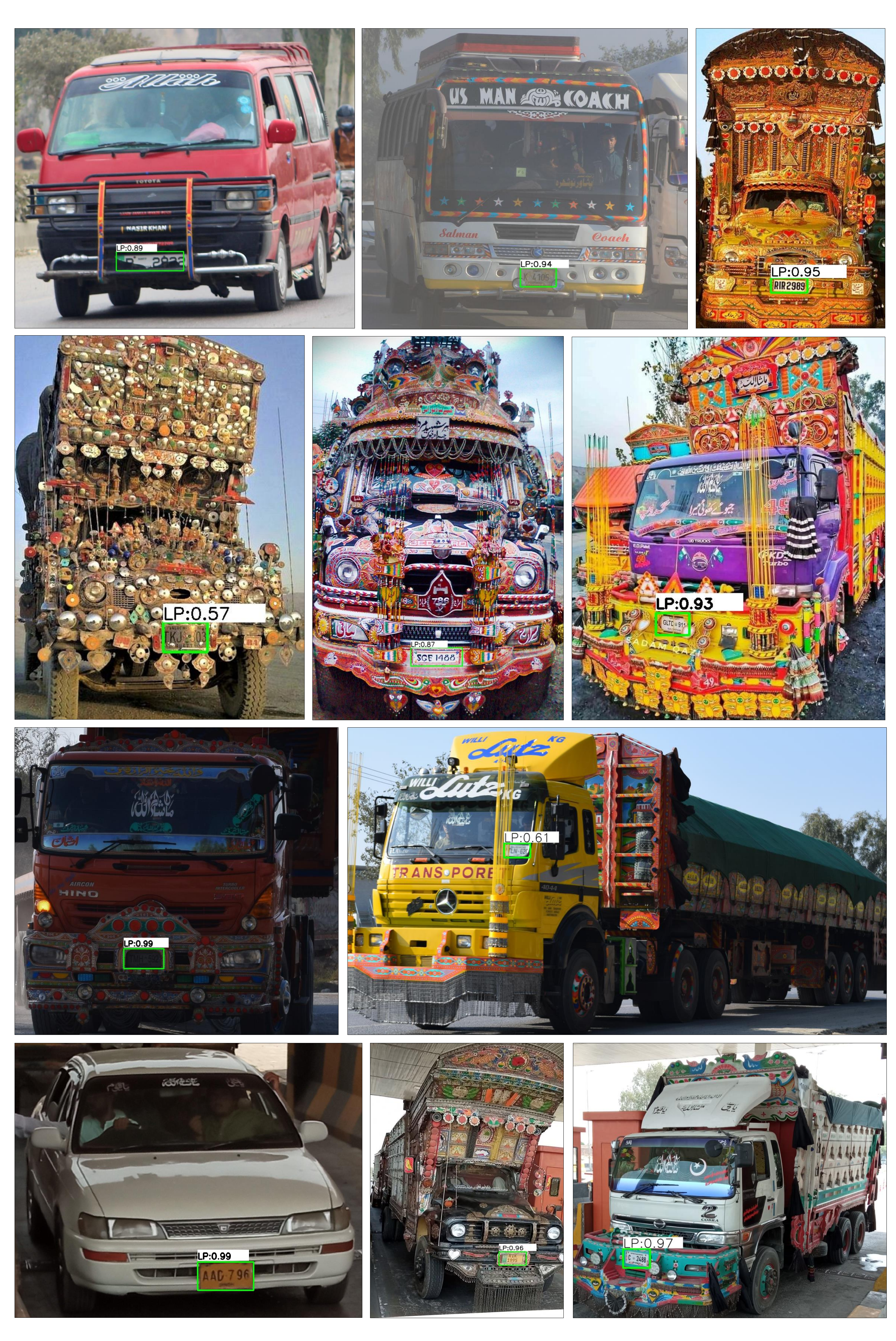}
    \caption{License Plate Detection result. The last row shows the detection results on the images taken at the toll plaza}
    \label{fig:11}
\end{figure}
This is due to several factors; for instance, the positions of the license plates on buses are more coherent than those on trucks. Secondly, the decorations again play a degrading role in license plate detections as trucks; specifically, type-1 trucks are more decorated than buses, cars, and vans. Third, most of the type-1 trucks do not display the official license plates, due to which their fonts and styles are not coherent, hence resulting in a lower detection rate. 
Some license plate detection results achieved by YOLOv4 are shown in Figure~\ref{fig:11}, where the last row shows the detections on a toll plaza. It can detect license plates with extreme background clutter, such as those of the heavily decorated trucks. Interestingly, damaged license plates and those displayed behind the vehicle's windshield are also accurately detected. Furthermore, it is also able to detect license plates in dusty conditions. The best overall mAP of 98.5\% is achieved by YOLOv4 followed by YOLOv3 and Tiny YOLOv4.

\subsection{License plate characters and digits recognition}
\begin{figure}[t]
   \centering
   \includegraphics[width=0.7\columnwidth]{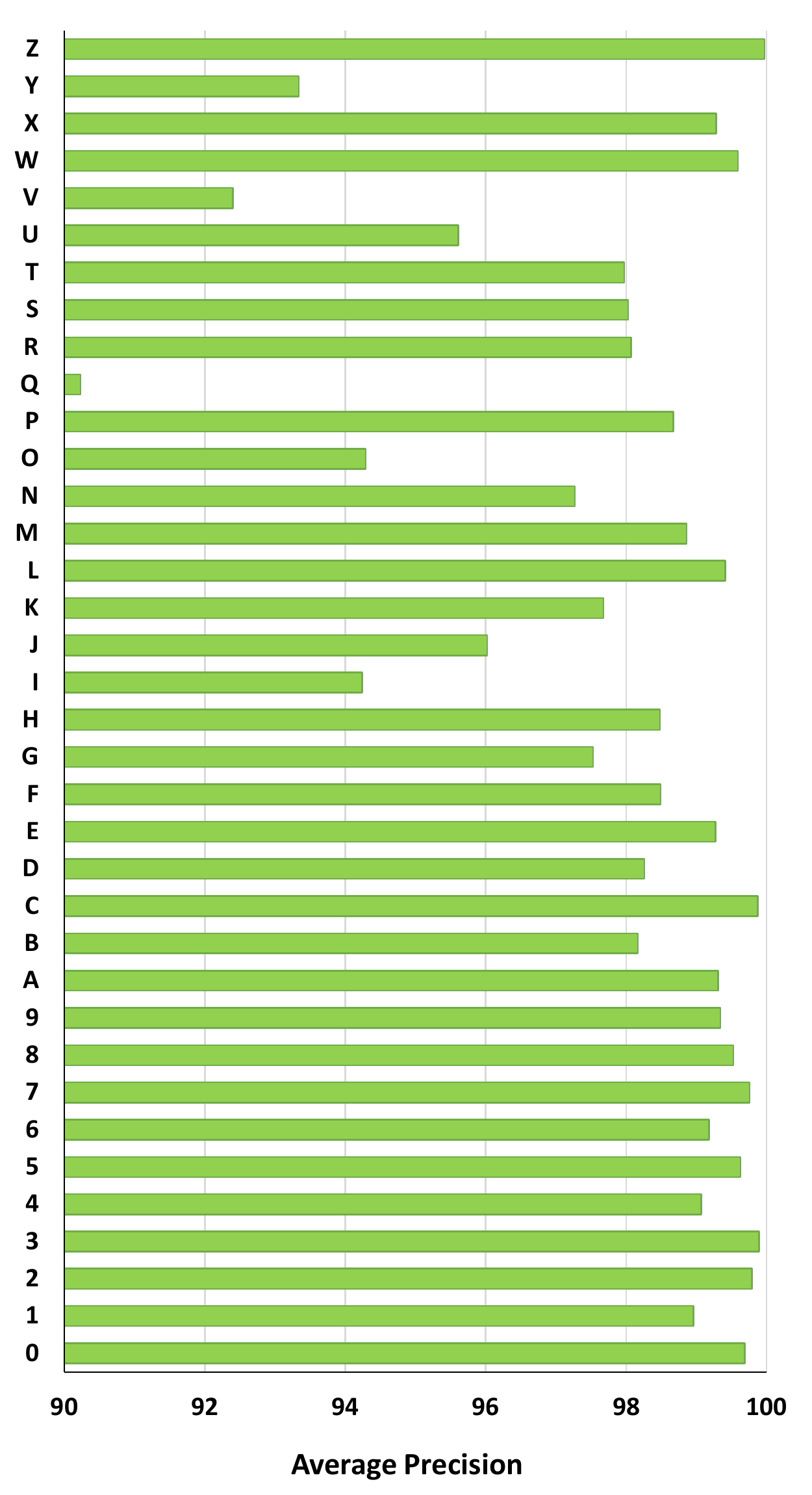}
   \caption{YOLOv4 average precision ($x$-axis) for each of the digits and characters classes ($y$-axis) }
    \label{fig:12}
\end{figure}

\begin{figure}[t]
    \centering
    \includegraphics[width=1\columnwidth]{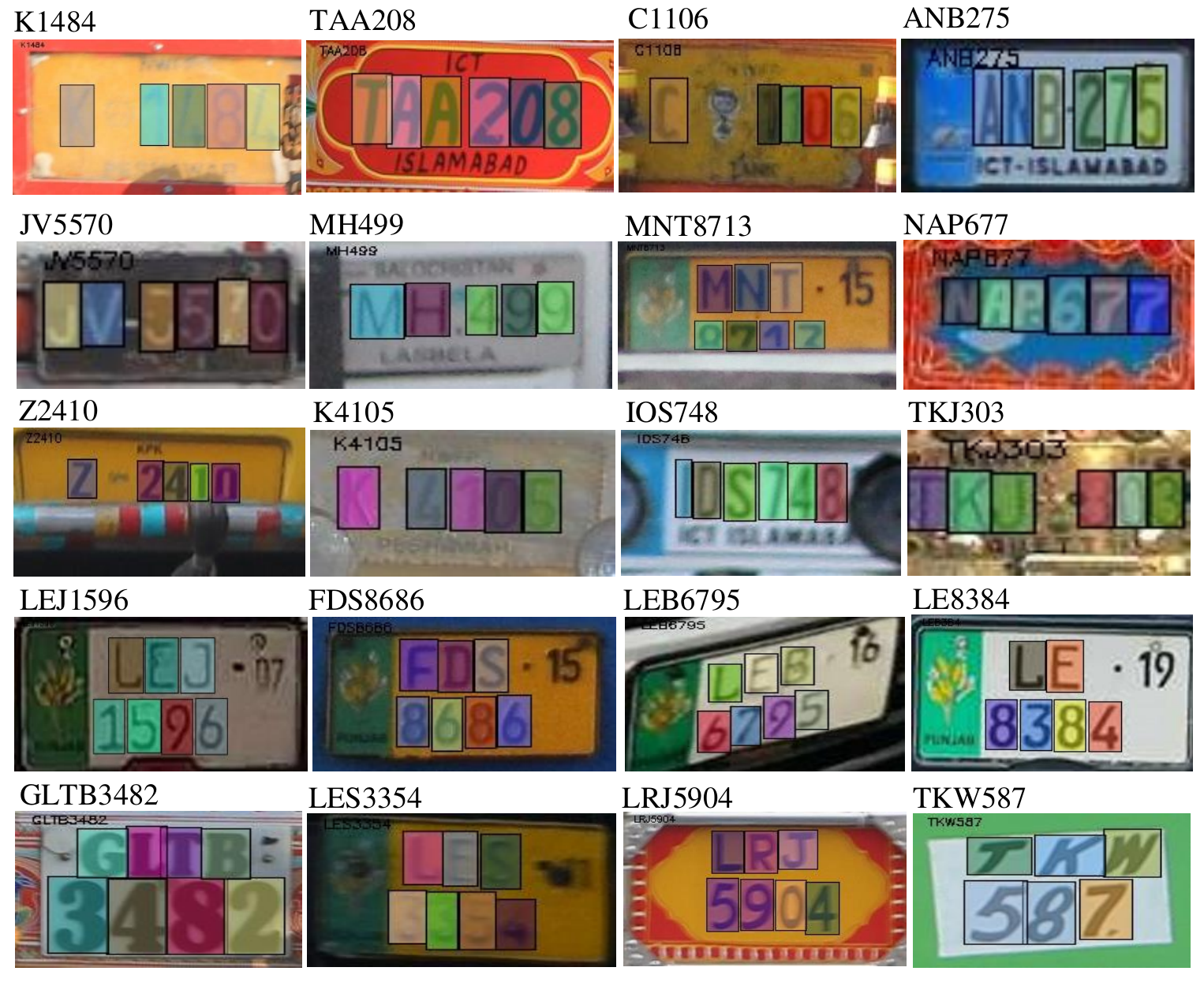}
    \caption{Character and digits recognition results on the license plate. The output of the framework is shown on the top of each image.}
    \label{fig:13}
\end{figure}

The recognition of characters and digits on Pakistani license plates is a challenging task due to the variations found in their fonts and the license plates layouts. As they are depicted on the license plates, their depictions are suffered from dust and mud, thus making their image-based recognition non-trivial. We train and test the third model named ``\textit{CR-Net}'' 5 times, where each time the dataset is randomly split. For each of the experiments, the number of training iterations is 10K due to the relatively higher number of classes in the \textit{lp-read dataset} which is 36. The results are given in Table~\ref{tab:char-recog}, whereas the average precision achieved by the architectures for each of the character's classes is shown in Figure~\ref{fig:12}. YOLOv4 performs the highest mAP of 98.3\%, followed by YOLOv3 and then Tiny YOLOv4. Figure~\ref{fig:13} shows some license plate character and digits prediction outputs. As can be seen, our model robustly and effectively tackles the different inconsistencies and variations present in the character dataset. Blurriness, angle shot, different font and font sizes, and resolution do not affect the performance and accuracy of our model to a large extent.

\begin{table}[t]
\caption{Character recognition results achieved by various architectures}
\label{tab:char-recog}
\begin{tabularx}{\columnwidth}{L{1.9cm} C{1.2cm} C{1.2cm} C{1.2cm} C{1.2cm}}
\toprule
\multicolumn{5}{c}{\textbf{Character Recognition}}                                                                                                                              \\ \midrule
                      & \multicolumn{1}{c}{\textbf{Precision}} & \multicolumn{1}{c}{\textbf{Recall}} & \multicolumn{1}{c}{\textbf{F1-score}} & \multicolumn{1}{c}{\textbf{mAP@0.5}} \\ \midrule

\textbf{Tiny YOLOv4}    & 94.0$\pm$2.0 & 94.4$\pm$2.2 & 94.4$\pm$1.5                             & 93.8$\pm$0.5              \\

\textbf{Tiny YOLOv3}    & 89.8$\pm$1.2 & 92.0$\pm$1.7 & 91.4$\pm$1.0                             & 92.4$\pm$1.0                       \\

\textbf{Faster-RCNN}    & 93.0$\pm$1.4 & 93.2$\pm$1.6  & 93.2$\pm$1.5                            & 93.0$\pm$1.0                          \\

\textbf{YOLOv2}         & 91.6$\pm$1.3 & 91.8$\pm$1.7 & 91.8$\pm$1.1                             & 92.8$\pm$0.7                        \\

\textbf{YOLOv3}         & 97.6$\pm$0.8 & 95.4$\pm$1.6 & 96.8$\pm$1.0                             & 97.0$\pm$0.4                          \\

\textbf{YOLOv4}         & 97.2$\pm$0.7 & 97.0$\pm$1.1 & 97.0$\pm$0.6                             & 98.3$\pm$0.5               \\ \bottomrule
\end{tabularx}
\end{table}
\subsection{End-to-End recognition}
Nonetheless, we aim for an end-end system for toll tax collection that can perform vehicle type and license plate recognition with the highest possible accuracy. For this purpose, we collect another test dataset where the images are taken at a toll tax plaza. This dataset contains 100 images per vehicle type, and we call it ``\textit{end-to-end dataset}''. We provide each of the images to the sequential pipeline formed by \textit{Vehicle-Net}, \textit{LP-Net} and \textit{CR-Net}. The backbone architecture used for each model is YOLOv4 due to its best performance on individual tasks. The end-to-end recognition is considered correct only if the vehicle type and license plate are correctly recognized. The results achieved for each of the vehicle types for an end-to-end recognition are shown in Table~\ref{tab:end-to-end}. As expected, cars and vans have the highest accuracy because they are not heavily decorated compared to trucks. Secondly, the positions of license plates on these vehicles are coherent. Lastly, on most of the cars and vans, the official computerized license plates are used, due to which there are fewer variations in their fonts and designs. As expected, both types of trucks have the least end-to-end recognition rates due to their heavy decorations and variations in license plates positions, fonts, and designs. 
\begin{table}[t]
\caption{End-to-end recognition rates for each vehicle type}
\label{tab:end-to-end}
\begin{center}
\begin{tabular}{l l}
\toprule
\textbf{Vehicle type} & \textbf{Recognition Rates} \\ \midrule
\textbf{Car}          & 99\%                       \\
\textbf{Van}          & 98\%                       \\
\textbf{Carry/Van}    & 98\%                       \\
\textbf{Bus}          & 97\%                       \\
\textbf{Type2-Truck}  & 97\%                       \\
\textbf{Type1-Truck}  & 96\%                       \\ \bottomrule
\end{tabular}
\end{center}
\end{table}
\subsection{Execution Time}
We aim to deploy the proposed framework on a Raspberry Pi to collect the toll tax on plazas. Due to this reason, the images acquired by the Pi camera should be efficiently processed for a vehicle type and license plate recognition in order to calculate the toll tax in real-time. In this regard, we perform experiments to estimate the inference time taken by our proposed framework with both YOLOv4 and Tiny YOLOv4 as its backbone architectures. The comparative time analysis is carried out on Nvidia GTX1060 GPU and Raspberry Pi, and results are shown in Table~\ref{tab:exec-time}. As expected, due to the lighter architecture, Tiny YOLOv4 takes less time on both GPU and Raspberry Pi than YOLOv4.
\begin{table}[t]
\caption{The computational time of YOLOv4 and Tiny YOLOv4 on GPU along with Raspberry Pi for all the three models}
\label{tab:exec-time}
\begin{center}
\begin{tabularx}{\columnwidth}{L{1.5cm} C{1.3cm} C{1.3cm} C{1.3cm} C{1.3cm}}
\toprule
\multirow{3}{*}{\textbf{}} & \multicolumn{4}{c}{\textbf{Execution time (sec)}}                                  \\ \cmidrule(l){2-5} 
                           & \multicolumn{2}{c}{\textbf{GPU}}       & \multicolumn{2}{c}{\textbf{Raspberry Pi}} \\ \cmidrule(l){2-5} 
                           & \textbf{YOLOv4} & \textbf{Tiny YOLOv4} & \textbf{YOLOv4}   & \textbf{Tiny YOLOv4}  \\ \midrule
\textbf{Vehicle-Net}       & 1.2             & 0.16                 & 7.1               & 0.95                  \\
\textbf{LP-Net}            & 0.9             & 0.14                 & 6.5               & 0.8                   \\
\textbf{CR-Net}            & 0.8             & 0.13                 & 5.0                 & 0.7                   \\ \bottomrule
\end{tabularx}
\end{center}
\end{table}
\subsection{Graphical user interface for real-time deployment}
\begin{figure*}[t!]
     \centering
    \includegraphics[width=1\textwidth]{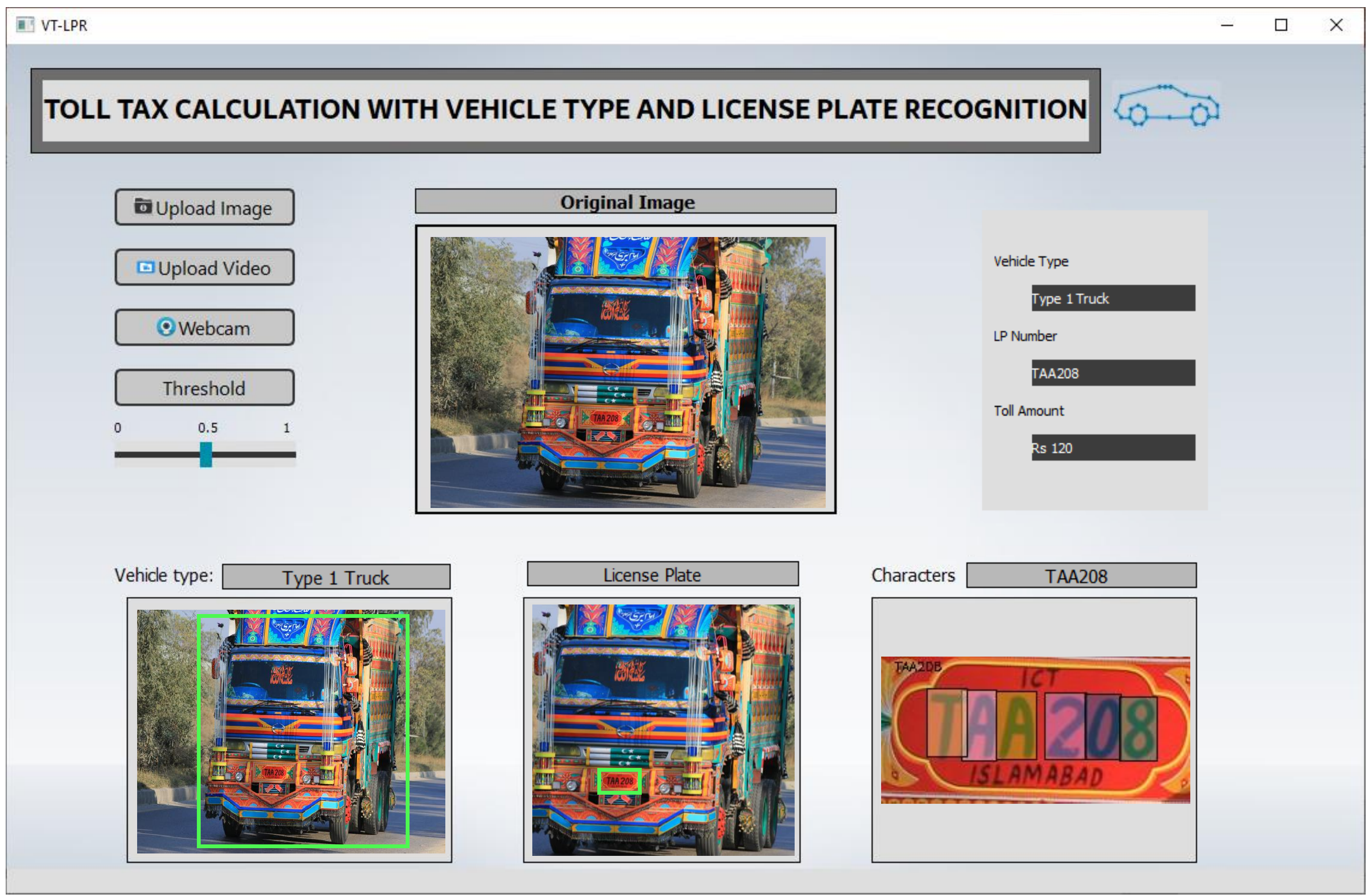}
     \caption{Graphical User Interface (GUI) for real time deployment of our proposed framework.}
     \label{fig:14}
\end{figure*}

Figure~\ref{fig:14} shows the user-friendly graphical interface of our proposed toll tax collection framework. We have given the flexibility of using both images and videos. The threshold is set to 0.5 by default that can be changed depending on the environment parameters. Lastly, outputs of all three steps \ie, vehicle recognition, license plate detection, and character recognition are shown pictorially and in text.
\section{Conclusion}
\label{sec:conclusion}
We present a framework for image-based automatic toll tax calculation of Pakistani vehicles, including cars, buses, carry vans, vans, and trucks. This framework can be instrumental in avoiding traffic jams on toll tax collection plazas and can prove an economical alternative to RFID-based systems. Our proposed framework is based on a three-step strategy involving vehicle detection and recognition in images, license plate detection on the detected vehicle, and finally, license plate digits and characters recognition. We evaluated variants of YOLO (v2, v3, and v4), their lighter versions (Tiny YOLOv3 and Tiny YOLOv4), and FasterRCNN for all the three steps on an image dataset of 10K images of Pakistani vehicles on roads and in toll plazas. Overall, YOLOv4 performed the best on all three steps individually and then combined on an end-to-end recognition that spans all three models sequentially. Its lighter version, \ie, Tiny YOLOv4, performed second best; however, it took less time than YOLOv4 on GPU and Raspberry Pi. 

Possible future directions of the current research are more focused on trucks. We get the least average precision on them due to the variations in their shapes and models. We also plan to evaluate more object detection frameworks to get better mAP on real-time images and videos.  
\section*{Acknowledgment}
Hafeez Anwar is supported by the Austrian Agency for International Cooperation in Education and Research (OeAD) under the Ernst Mach follow-up grant program.

\bibliographystyle{IEEEtran}
\bibliography{IEEEabrv,mybib}

\end{document}